\documentclass{article}
\usepackage{amsmath,amssymb,amsthm,graphicx,color,bm}
\usepackage{enumerate}
\usepackage{latexsym,natbib}
\usepackage{bbm}
\usepackage{times}
 \renewcommand{\And}{\wedge}
  \newtheorem{thm}{Theorem}[section]
   \newtheorem{prop}[thm]{Proposition}

\newtheorem{cor}[thm]{Corollary}

\newcommand{\xsection}[1]{\section{#1}}

  \newcommand{\beqa}{\begin{eqnarray}}
  \newcommand{\eeqa}{\end{eqnarray}}
  \newcommand{\beqas}{\begin{eqnarray*}}
  \newcommand{\eeqas}{\end{eqnarray*}}
  \newcommand{\benum}{\begin{enumerate}[{\rm (i)}]\itemsep-4pt}
\newcommand{\eenum}{\end{enumerate}}
 \newcommand{\C}{{\bf C}}
 \newcommand{\R}{{\mathbbm{R}}}
\newcommand{\V}{V}
 \newcommand{\al}{\alpha}
  \newcommand{\be}{\beta}
  \newcommand{\et}{\eta}
 \newcommand{\mb}{\mbox}
  \newcommand{\om}{\omega}
  \newcommand{\ph}{\phi}
\newcommand{\ps}{\psi}
  \newcommand{\vp}{\varphi}
  \newcommand{\De}{\Delta}
 \newcommand{\Ga}{\Gamma}
  \newcommand{\IFF}{\Leftrightarrow}
  \newcommand{\Iff}{\leftrightarrow}
  \newcommand{\Inf}{\bigwedge}
  \newcommand{\Not}{\neg}
 \newcommand{\Or}{\vee}
  \newcommand{\Sup}{\bigvee}
 \newcommand{\THEN}{\Rightarrow}
  \newcommand{\Then}{\rightarrow}
  \newcommand{\dom}{\mathop{\rm dom}}
\newcommand{\bra}[1]{\langle#1|}
\newcommand{\ket}[1]{|#1\rangle}
\newcommand{\ketbra}[1]{\ket{#1}\bra{#1}}
\newcommand{\bracket}[1]{\langle#1\rangle}
  \newcommand{\rank}{\mbox{\rm rank}}
  \newcommand{\ran}{\mbox{\rm ran}}
  \newcommand{\bS}{{\bf S}}
 \newcommand{\cA}{{\cal A}}
  \newcommand{\cB}{{\cal B}}
  \newcommand{\cC}{{\cal C}}
  \newcommand{\cF}{{\cal F}}
  \newcommand{\cH}{{\cal H}}
  \newcommand{\cL}{{\cal L}}
  \newcommand{\cM}{{\cal M}}
  \newcommand{\cO}{{\cal O}}
  \newcommand{\cP}{{\cal P}}
  \newcommand{\cQ}{{\cal Q}}
  \newcommand{\cS}{{\cal S}}
 \newcommand{\cU}{{\cal U}}
  \newcommand{\tP}{\tilde{P}}
  \newcommand{\tQ}{\tilde{Q}}
\newcommand{\VQ}{V^{(\cQ)}}
\newcommand{\VL}{V^{(\cL)}}
\newcommand{\VB}{V^{(\cB)}}
\newcommand{\LL}{\bm{L}}
\renewcommand{\inf}{\bigwedge}
\renewcommand{\sup}{\bigvee}
\newcommand{\forces}{\vdash\!\!\!\vdash}
\newcommand{\val}[1]{[\![#1]\!]}
\newcommand{\valo}[1]{[\![#1]\!]_{o}}
\newcommand{\p}{{}^{\perp}}
\newcommand{\ck}[1]{\check{#1}}
\newcommand{\id}{{\rm id}}
\newcommand{\commutes}{\ {}^{|}\!\!\!{}_{\circ}\ }
\newcommand{\com}{{\rm com}}
\newcommand{\hvp}{{\hat{\vp}}}

\title{ORTHOMODULAR-VALUED MODELS FOR QUANTUM SET THEORY%
}
 \author{MASANAO OZAWA\\
 Graduate School of Informatics, Nagoya University}
 \date{}
\begin{document}
\maketitle

\begin{abstract}
In 1981, Takeuti  introduced quantum set theory by constructing a model of set theory 
based on quantum logic represented by the lattice of closed linear subspaces 
of a Hilbert space in a manner analogous to Boolean-valued models of set theory,
and showed that appropriate counterparts of the axioms of Zermelo-Fraenkel set
theory with the axiom of choice (ZFC) hold in the model.
In this paper, we aim at unifying Takeuti's model with Boolean-valued models
by constructing models based on general complete orthomodular lattices,
and generalizing the transfer principle in Boolean-valued models, 
which asserts that every theorem in ZFC set theory
holds in the models, to a general form holding in every orthomodular-valued model.  
One of the central problems in this program is the well-known arbitrariness in choosing 
a binary operation for implication.  To clarify what properties are required 
to obtain the generalized transfer principle, we introduce a class of 
binary operations extending the implication on Boolean logic, 
called generalized implications, including even non-polynomially definable
operations.
We study the properties of those operations in detail and show that all of them admit
the generalized transfer principle.  Moreover, we determine all the polynomially 
definable operations for which the generalized transfer principle holds.
This result allows us to abandon the Sasaki arrow originally assumed 
for Takeuti's model and leads to a much more flexible approach to quantum set theory.
\end{abstract}

\xsection{Introduction.}
The notion of sets has been considerably extended since \cite{Coh63,Coh66} developed
the method of forcing for the independence proof of the continuum hypothesis.
After Cohen's work, the forcing subsequently became a central method 
in axiomatic set theory and was incorporated into various notions in 
mathematics, in particular, 
the notion of sheaves \citep{FS79} and sets in nonstandard logics,
such as the Boolean-valued set theory 
reformulating the method of forcing  \citep{SS67},
topos \citep{Joh77}, and intuitionistic set theory \citep{Gra79}. 
Quantum set theory was introduced by 
\cite{Ta81} as a successor of these attempts, extending the notion of sets
to be based on quantum logic introduced by \cite{BvN36}.

Let $\cB$ be a complete Boolean algebra.
\cite{SS67} introduced the Boolean-valued model $\VB$ for set theory
with $\cB$-valued truth value assignment $\val{\vp}$ for formulas $\vp$ 
of set theory and showed the following fundamental theorem 
for Boolean-valued models $\VB$ 
\citep[Theorem 1.33]{Bel05}.

{\bf Boolean Transfer Principle.}
{\em For any formula $\vp(x_1,\ldots,x_n)$ provable in ZFC set theory,
the $\cB$-valued truth value $\val{\vp(u_1,\ldots,u_n)}$  satisfies
\beqas
\val{\vp(u_1,\ldots,u_n)}=1
\eeqas
for any $u_1,\ldots,u_n\in\VB$. }

For a given sentence $\ph$ of ZFC set theory, if we can construct a complete
Boolean algebra $\cB$ such that $\val{\ph}<1$ in $\VB$, then we can conclude 
that $\ph$
is not provable in ZFC.
Let CH denote the continuum hypothesis. 
It is shown  that if $\cB$ is the complete Boolean algebra 
of the Borel subsets modulo the null sets of the product measure space 
$\{0,1\}^{\aleph_0\times I}$, where $I>2^{\aleph_0}$, then $\val{\rm CH}=0$,
and the independence of CH from axioms of ZFC follows  
\citep[Theorem 19.7]{TZ73}.

Based on the standard quantum logic represented by 
the lattice $\cQ$ of closed subspaces of a Hilbert space $\cH$,
\cite{Ta81} constructed the universe $\VQ$ of 
set theory with $\cQ$-valued truth value assignment $\val{\vp}$
for formulas $\vp$ of set theory 
in a manner similar to the Boolean-valued universe $\VB$ 
based on a complete Boolean algebra $\cB$.
As one of the promising aspects,
 \cite{Ta81} showed that the real numbers 
in $\VQ$ are in one-to-one correspondence with
the self-adjoint operators on the Hilbert space $\cH$,
or equivalently the observables of the quantum system described by $\cH$.
As a difficult aspect, it was also revealed that quantum set theory is so
irregular that the transitivity law and the substitution rule for equality 
do not generally hold without modification.
To control the irregularity, \cite{Ta81} introduced the commutator 
$\com(u_1,\ldots,u_n)$ of elements ($\cQ$-valued sets) $u_1,\ldots,u_n$ of 
the universe $\VQ$ and showed that each axiom of ZFC can be modified 
through commutators to be a sentence valid in $\VQ$.

In a preceding paper \citep{07TPQ}, the present author further advanced
Takeuti's use of the commutator and established the following general principle:

{\bf Quantum Transfer Principle.}
{\em The $\cQ$-valued truth value $\val{\vp(u_1,\ldots,u_n)}$ of any 
$\De_0$-formula $\vp(x_1,\ldots,x_n)$ provable in ZFC set theory 
satisfies
\beqas
\val{\vp(u_1,\ldots,u_n)}\ge\com(u_1,\ldots,u_n)
\eeqas
for any $u_1,\ldots,u_n\in\VQ$. }

The Quantum Transfer Principle is obviously a quantum counter part of the 
Boolean Transfer Principle.
To deepen the Quantum Transfer Principle we consider the following
two problems:
\benum
\item
Unify Takeuti's models and Boolean valued models providing
the same footing for the Quantum Transfer Principle and the Boolean Transfer Principle.
\item
Determine the binary operations that can be used for 
implication in order for the model $\VQ$ 
to satisfy the Quantum Transfer Principle?
\eenum
Problem 1 was partially solved in the preceding paper \citep{07TPQ}, in which 
Takeuti's model $\VQ$ was generalized to the logic represented 
by the complete orthomodular lattice $\cQ=\cP(\cM)$ of projections 
in a von Neumann algebra $\cM$ on a Hilbert space $\cH$,
and the Quantum Transfer Principle was 
actually proved under this general formulation. 
This generalization enables us to apply quantum set theory to algebraic 
quantum field theory \citep{Ara00} as well as classical mechanics in
a unified framework.    
However, from a set theoretical point of view, this framework is not broad enough, 
as the class of complete Boolean subalgebras $\cB$ in $\cQ=\cP(\cM)$ 
excludes set-theoretically interesting Boolean algebras such as cardinal 
collapsing algebras.  This follows from the fact that every 
complete Boolean subalgebra $\cB$ of $\cQ=\cP(\cM)$  satisfies the
local countable chain condition \citep[p.~118]{Ber72} so that 
the cardinals are absolute in $\VB$ \citep[p.~50]{Bel05}.
In this paper, we generalize Takeuti's model to the class of 
complete orthomodular lattices, 
which includes  all the complete Boolean algebras,
as well as
all the projection lattices of von Neumann algebras.

Problem 2 relates to a longstanding 
problem in quantum logic concerning the
arbitrariness in choosing a binary operation for implication.
It is known that there are exactly six ortholattice polynomials that reduces to
the classical implication $P\Then Q=\Not P\Or Q$ on Boolean algebras \citep{Kot67}.
Among them, the majority favor the Sasaki arrow $P\Then Q=P^{\perp}\Or(P\And Q)$ \citep{Urq83}.
In fact, following \cite{Ta81}, the preceding work \citep{07TPQ} 
adopted the Sasaki arrow for implication to establish the Quantum Transfer Principle.
Here, to treat the most general class of binary operations,
we introduce the class of generalized implications in complete
orthomodular lattices characterized by simple conditions and 
including the above-mentioed six polynomials 
as well as continuously many non-polynomial binary operations,
which are defined through non-polynomial binary operations 
introduced in the standard quantum logic by \cite{Ta81}.
We introduce the universe $\VQ$ of sets based on a complete orthomodular lattice 
$\cQ$ with a generalized implication, 
and show that the Quantum Transfer Principle always holds in this 
general formulation.  We also determine all the polynomially 
definable operations for which the Quantum Transfer Principle holds.
This result allows us to abandon the Sasaki arrow assumed in previous
formulations and leads to a much more flexible approach to quantum set theory.
In this general formulation, the Quantum and 
Boolean Transfer Principles can be treated on the same footing.
Moreover, we show that the Boolean Transfer Principle holds if and only if $\cQ$ is a Boolean
algebra.

This paper is organized as follows.
\ref{se:Preliminary}
collects basic properties of complete orthomodular 
lattices.
In \ref{se:GIIQL}, we introduce generalized implications in complete
orthomodular lattices and show their basic properties.
In \ref{se:NPIIQL}, by using non-polynomial binary operations
introduced by \cite{Ta81},
we show that there are continuously many 
different generalized implications that are not polynomially definable
even in the standard quantum logic, and provide their basic properties.
\ref{se:UOQS} introduces 
the universe of sets based on a complete orthomodular lattice 
with a generalized implication, and show some basic properties.
In  \ref{se:TPIQST}, we prove the Quantum Transfer Principle
for any complete orthomodular lattice with a generalized implication.
We also determine all the polynomially definable binary operations
for which the Quantum Transfer Principle holds.
Moreover, we show that the Boolean Transfer Principle holds 
if and only if $\cQ$ is a Boolean algebra.

\xsection{Preliminaries.}
\label{se:Preliminary}
\subsection{Quantum logic.}
\label{se:QL}

A {\em complete orthomodular lattice}  is a complete
lattice $\cQ$ with an {\em orthocomplementation},
a unary operation $\perp$ on $\cQ$ satisfying

(C1) if $P \le Q$ then $Q^{\perp}\le P^{\perp}$,

(C2) $P^{\perp\perp}=P$,

(C3)  $P\Or P^{\perp}=1$ and $P\And P^{\perp}=0$,

\noindent
where $0=\Inf\cQ$ and $1=\Sup\cQ$,
that  satisfies the {\em orthomodular law}:

(OM) if $P\le Q$ then $P\Or(P^{\perp}\And Q)=Q$.

\noindent\sloppy
In this paper, any complete orthomodular lattice is called a {\em logic}.
We refer the reader to 
\cite{Kal83} for a standard text on
orthomodular lattices.
In what follows, 
$P,Q,P_\al,\ldots$denote general elements of a logic $\cQ$. 

The orthomodular law weakens the distributive law, so that
any complete Boolean algebra is a logic.
The projection lattice $\cP(\cM)$ of 
a von Neumann algebra $\cM$ on a Hilbert space
$\cH$ is a logic \cite[p.~69]{Kal83}.  
The lattice $\cC(\cH)$ of closed subspaces of a Hilbert space 
$\cH$ with the operation of orthogonal complementation
is most typically a logic, the so-called {\em standard quantum logic}
on $\cH$,  and is isomorphic to $\cQ(\cH)=\cP({\rm B}(\cH))$, 
the projection lattice of the algebra 
${\rm B}(\cH)$ of bounded operators on $\cH$ \cite[p.~65]{Kal83}.

A non-empty subset of a logic $\cQ$ is called a {\em sublattice} if
it is closed under $\And $ and $\Or$.
A sublattice is called a {\em subalgebra} if it is further closed under 
$\perp$.
A sublattice or a subalgebra $\cA$ of $\cQ$ is said to be {\em complete} if it has
the supremum and the infimum in $\cQ$ of an arbitrary subset of $\cA$.
For any subset $\cA$ of $\cQ$, the sublattice generated by $\cA$ is
denoted by $[\cA]_0$, the complete sublattice generated 
by $\cA$ is denoted by $[\cA]$,
the subalgebra generated by $\cA$ is denoted by
$\Ga_0\cA$, and the complete subalgebra 
generated by $\cA$ is denoted by $\Ga\cA$, 

We say that $P$ and $Q$ in a logic $\cQ$
{\em commute}, in  symbols
$P\commutes Q$, if  $P=(P\And Q)\Or(P\And
Q^{\perp})$. All the relations $P\commutes Q$,
$Q\commutes P$,
$P^{\perp}\commutes Q$, $P\commutes Q^{\perp}$,
and $P^{\perp}\commutes Q^{\perp}$ are equivalent.
The distributive law does not hold in general, but
the following useful propositions  hold \cite[pp.~24--25]{Kal83}.

\begin{prop}\label{th:distributivity}
If $P_1,P_2\commutes
Q$, then the sublattice generated by $P_1,P_2,Q$ is
distributive.
\end{prop}

\begin{prop}\label{th:logic}
If $P_{\al}\commutes Q$ for all $\al$, then
$\Sup_{\al}P_{\al}\commutes Q$, 
$\Inf_{\al}P_{\al}\commutes Q$,
$Q \And (\Sup_{\al}P_{\al})=\Sup_{\al}(Q\And
P_{\al})$,
and 
$Q \Or (\Inf_{\al}P_{\al})=\Inf_{\al}(Q\Or
P_{\al})$,
\end{prop}

When applying a distributive law under the assumption of 
Proposition \ref{th:distributivity}, we shall say that we are
{\em focusing} on $Q$.
From Proposition \ref{th:distributivity}, a logic $\cQ$ is a Boolean
algebra if and only if $P\commutes Q$  for all $P,Q\in\cQ$.

For any subset $\cA\subseteq\cQ$,
we denote by $\cA^{!}$ the {\em commutant} 
of $\cA$ in $\cQ$ \cite[p.~23]{Kal83}, i.e., 
$$
\cA^{!}=
\{P\in\cQ\mid P\commutes Q \mbox{ for all }
Q\in\cA\}.
$$
Then $\cA^{!}$ is a complete orthomodular
sublattice of 
$\cQ$, i.e., $\Inf \cS,\Sup \cS,
P^{\perp}\in\cA^{!}$ for any $\cS\subseteq\cA^{!}$ and
$P\in\cA^{!}$.
A {\em sublogic} of $\cQ$ is a subset $\cA$ of
$\cQ$ satisfying $\cA=\cA^{!!}$. 
Thus, any sublogic of $\cQ$ is a complete subalgebra of $\cQ$.
For the case where $\cQ=\cQ(\cH)$ for a Hilbert space 
$\cH$,
a sublogic is characterized as the lattice of projections in
a von Neumann algebra acting on $\cH$ \citep{07TPQ}.
For any subset $\cA\subseteq\cQ$, the smallest 
logic including $\cA$ is 
$\cA^{!!}$ called the  {\em sublogic generated by
$\cA$}.
We have $\cA\subseteq [\cA] \subseteq \Ga \cA\subseteq
\cA^{!!}$.
Then it is easy to see that subset 
 $\cA$ is a Boolean sublogic, or equivalently 
 a distributive sublogic, if and only if 
$\cA=\cA^{!!}\subseteq\cA^{!}$.
If $\cA\subseteq\cA^{!}$, the subset
$\cA^{!!}$ is the smallest Boolean sublogic including
$\cA$.
A subset $\cA$ is a maximal Boolean sublogic  if
and only if $\cA=\cA^{!}$.
By Zorn's lemma, for every subset $\cA$ consisting of
mutually commuting elements, there is a maximal Boolean
sublogic including $\cA$.

\subsection{Commutators.}
\label{se:CIQL}

Let $\cQ$ be a logic.
\cite{Mar70} has introduced the commutator $\com(P,Q)$ 
of two elements $P$ and $Q$ of $\cQ$ by 
\beqas
\com(P,Q)=(P\And Q)\Or(P\And Q\p)\Or(P\p\And Q)\Or(P\p\And Q\p).
\eeqas
\cite{BK73} have generalized this notion 
to finite subsets of $\cQ$ by 
\beqas
\com(\cF)=\Sup_{\al:\cF\to\{\id,\perp\}}\Inf_{P\in\cF}P^{\al(P)}
\eeqas
for all $\cF\in\cP_{\om}(\cQ)$,
where $\cP_{\om}(\cQ)$ stands for the set of finite subsets of $\cQ$, and
$\{\id,\perp\}$ stands for the set consisting of the identity operation $\id$ and the 
orthocomplementation~$\perp$.
Generalizing this notion to arbitrary subsets $\cA$ of $\cQ$, 
\cite{Ta81} defined
$\com(\cA)$ by
\beqas
\com(\cA)&=&\Sup T(\cA),\\
T(\cA)&=&\{E\in\cA^{!} \mid P_{1}\And E\commutes P_{2}\And E
\mb{ for all }P_{1},P_{2}\in\cA\}
\eeqas
for any $\cA\in\cP(\cQ)$, where $\cP(\cQ)$ stands for the power set of $\cQ$,
and showed that $\com(\cA)\in T(\cA)$.
Subsequently, 
\cite{Pul85} showed:
\begin{thm}
For any subset $\cA$ of a logic $\cQ$, we have
\begin{enumerate}[(ii)]
\item[\rm (i)] $\com(\cA)=\Inf\{\com(\cF)\mid \cF\in\cP_{\om}(\cA)\},$
\item[\rm (ii)]  $\com(\cA)=\Inf\{\com(P,Q)\mid P,Q\in \Ga_0(\cA)\}$.
\end{enumerate}
\end{thm}

Let $\cA\subseteq\cQ$.
Denote by $L(\cA)$ the sublogic generated by $\cA$, i.e., 
$L(\cA)=\cA^{!!}$,
and by $Z(\cA)$ the center of $L(\cA)$, i.e., $Z(\cA)=\cA^{!}\cap\cA^{!!}$.
A {\em subcommutator} of $\cA$ is  any $E\in Z(\cA)$ 
such that $P_1\And E\commutes P_2\And E$ for all $P_1,P_2\in\cA$.
Denote by $T_0(\cA)$ the set of subcommutators of  $\cA$, i.e., 
\beqa
T_0(\cA)=\{E\in Z(\cA)\mid P_{1}\And E\commutes P_{2}\And E
\mb{ for all }P_{1},P_{2}\in\cA\}.
\eeqa

For any $P,Q\in\cQ$, the {\em interval} $[P,Q]$ is the set of
all $X\in\cQ$ such that $P\le X\le Q$.
For any $\cA\subseteq \cQ$ and $P,Q\in\cA$,
we write $[P,Q]_{\cA}=[P,Q]\cap\cA$.
Then the following theorem holds \citep{16A2}.
\begin{thm}\label{th:equivalence_com}
For any subset  $\cA$  of a logic $\cQ$, the following hold.
\benum
\item[\rm (i)] $T_0(\cA)=\{E\in Z(\cA)\mid [0,E]_{\cA}\subseteq Z(\cA)\}$.
\item[\rm (ii)] $\Sup T_0(\cA)$ is the maximum subcommutator of $\cA$, 
i.e., $\Sup T_0(\cA)\in T_0(\cA)$.
\item[\rm (iii)] $T_0(\cA)=[0,\Sup T_0(\cA)]_{L(\cA)}$.
\item[\rm (iv)] $\com(\cA)=\Sup T_0(\cA)$.
\eenum
\end{thm}

The following proposition will be useful in later discussions \citep{16A2}.

\begin{thm}\label{th:maximal_Boolean}
Let $\cB$ be a maximal Boolean sublogic of a logic $\cQ$ and $\cA$ a
subset of $\cQ$ including $\cB$, i.e., $\cB\subseteq\cA\subseteq\cQ$.
Then  $\com(\cA)\in\cB$ and $[0,\com(\cA)]_{L(\cA)}\subset \cB$.
\end{thm}

The following theorem clarifies the significance of commutators \citep{16A2}.

\begin{thm}\label{th:absoluteness_commutator}
Let $\cA$ be a subset of a logic $\cQ$.
Then  $L(\cA)$ is isomorphic to the direct product of the complete Boolean algebra
$[0,\com(\cA)]_{L(\cA)}$ and the complete orthomodular lattice
$[0,\com(\cA)^\perp]_{L(\cA)}$ without non-trivial Boolean factor.
\end{thm}
\sloppy

We refer the reader to 
\cite{Pul85} and 
\cite{Che89} for further results about commutators
in orthomodular lattices.

\xsection{Generalized implications in quantum logic.}
\label{se:GIIQL}

In classical logic, the implication connective 
$\Then$ is defined by negation $\perp$ and
disjunction $\Or$ as $P\Then Q=P^{\perp}\Or Q$.
In quantum logic, several counterparts have been proposed.  
\cite{Har81} proposed the following
requirements for the implication connective.
\benum
\item[(E)]    $P\Then Q=1$ if and only if $P\le Q$ for all $P,Q\in\cQ$.
\item[(MP)] $P             \And (P\Then Q) \le Q$ for all $P,Q\in\cQ$.
\item[(MT)] $Q^{\perp}\And (P\Then Q) \le P^{\perp}$ for all $P,Q\in\cQ$.
\item[(NG)] $P\And Q\p\le(P\Then Q)^\perp$ for all $P,Q\in\cQ$.
\item[(LB)] If $P\commutes Q$, then 
$P\Then Q=P^{\perp}\Or Q$ for all $P,Q\in\cQ$.
\eenum
The work of 
\cite{Kot67} can be applied 
to the problem as to what
ortholattice-polynomials $P\Then Q$
satisfy the above conditions; see also \cite{Har81} and \cite{Kal83}.
There are exactly six two-variable ortholattice-polynomials satisfying (LB),
defined as follows.

(0) $P\Then_0 Q=(P\p\And Q\p)\Or(P\p\And Q)\Or(P\And Q)$.

(1) $P\Then_1 Q=(P\p\And Q\p)\Or(P\p \And Q)\Or(P\And(P\p\Or Q))$.

(2) $P\Then_2 Q=(P\p\And Q\p)\Or Q$.

(3) $P\Then_3 Q=P\p \Or(P\And Q)$.

(4) $P\Then_4 Q=((P\p\Or Q)\And Q\p)\Or(P\p \And Q)\Or(P\And Q)$.

(5) $P\Then_5 Q=P\p\Or Q$.

It is also verified that requirement (E) is satisfied by $\Then_j$ for $j$=0,\ldots,4
and that all requirements (E), (MP), (MT), (NG), and (LB) are satisfied by
$\Then_j$ for $j$=0,2,3.

We call 
$\Then_0$ the {\em minimum implication},
$\Then_2$ the {\em contrapositive Sasaki arrow},
$\Then_3$ the {\em Sasaki arrow}, and
$\Then_5$ the {\em maximum implication}.
So far we have no general
agreement on the choice from the above,  although the
majority view favors the Sasaki
arrow \citep{Urq83}.

As defined later in \ref{se:UOQS}, the truth values 
$\val{u\in v}$ and $\val{u=v}$ 
of atomic formulas in quantum set theory
depend crucially on the definition of 
implication connective.
\cite{Ta81} and 
the present author
\citep{07TPQ} previously chose the Sasaki arrow for this purpose.
However, there are several reasons for investigating wider choices of 
implication connective.
To mention one, 
consider De Morgan's law for bounded
quantifiers in set theory:
$$
\val{\Not (\exists x\in u)\vp(x)}=\val{(\forall x\in u)\Not\vp(x)}.
$$
The validity of this fundamental law depends on the choice of 
implication connective $\Then$, since the right-hand side is determined by
$$
\val{(\forall x\in u)\Not\vp(x)}=\Inf_{x\in\dom(u)}
u(x)\Then \val{\vp(x)}^{\perp},
$$
whereas the left-hand side is determined by the original lattice operations as
$$
\val{\Not(\exists x\in u)\vp(x)}=\left(\Sup_{x\in\dom(u)}
u(x)\And \val{\vp(x)}\right)^\perp.
$$
Remarkably, our previous choice, the Sasaki arrow, does not satisfy
this law, while only the maximum implication satisfies it.
Thus, we have at least one logical principle that favors the maximum 
implication which has been rather excluded 
because of its failure 
in satisfying (E), (MP), or (MT).
In this paper, we develop a quantum set theory based on a very general 
choice of implication to answer the question what properties 
of the implication ensure the transfer principle for quantum set theory.

A binary operation $\Then$ on a logic $\cQ$ is called a {\em generalized implication}
if the following conditions hold.

(I1)  $P\Then Q\in\{P,Q\}^{!!}$ for all $P,Q\in\cQ$.

(I2) $(P\Then Q)\And E
=[(P\And E)\Then(Q\And E)]\And E$ if $P, Q\commutes E$ for all $P,Q,E\in\cQ$.

(LB) If $P\commutes Q$, then $P\Then Q=P^{\perp}\Or Q$ for all $P,Q\in\cQ$.

We shall show that properties (I1), (I2), and (LB) suffice to ensure that the Quantum
Transfer Principle holds.  It is interesting to see that any polynomially definable 
binary operation has properties (I1)--(I2) as shown below.
Thus, the Quantum Transfer Principle holds for a polynomially definable 
implication if and only if it satisfies (LB), so that it is
exactly one of the six implications $\Then_j$ for $j=0,\ldots,5$.
Examples of non-polynomially definable generalized implications will be
given in \ref{se:NPIIQL}.  
They require (I1) instead of  $P\Then Q\in \Ga_0\{P,Q\}$.
They are derived by  Takeuti's 
non-polynomially definable operation introduce in \citep{Ta81},
for which \cite{Ta81} wrote ``We believe that we have to study this type of
new operation in order to see the whole picture of quantum set theory 
including its strange aspects''.

\begin{prop}
\label{th:restriction_property}
For any two-variable ortholattice polynomial  $f$ on a logic $\cQ$,
we have the following.
\benum
\item[{\rm (i)}] $f(P,Q)\in\{P,Q\}^{!!}$ for all $P,Q\in\cQ$.
\item[{\rm (ii)}] $f(P,Q)\And E=f(P\And E,Q\And E)\And E$ if $P, Q\commutes E$ for all $P,Q,E\in\cQ$.
\eenum
\end{prop}
\begin{proof}
Since $f(P,Q)\in\Ga_0\{P,G\}\subseteq\{P,Q\}^{!!}$,
statement (i) follows.
The proof of (ii) is carried out by induction on the complexity of the polynomial $f(P,Q)$.
First, note that from $P,Q\commutes E$ we have $g(P,Q)\commutes E$
for any two-variable polynomial $g$.  If $f(P,Q)=P$ or $f(P,Q)=Q$, 
assertion (ii) holds obviously.  If $f(P,Q)=g_1(P,Q)\And g_2(P,Q)$
with two-variable polynomials $g_1,g_2$, the assertion holds from associativity.
Suppose that  $f(P,Q)=g_1(P,Q)\Or g_2(P,Q)$
with two-variable polynomials $g_1,g_2$.
Since $g_1(P,Q), g_2(P,Q)\commutes E$, the assertion follows from the 
distributive law focusing on $E$.
Suppose $f(P,Q)=g(P,Q)^\perp$ with a two-variable polynomial $g$.  
For the case where $g$ is atomic, the assertion follows; for instance, if $g(P,Q)=P$, 
we have 
$f(P\And E,Q\And E)\And E=
(P\And E)^\perp\And E=
(P^\perp\Or E^\perp)\And E=P^\perp\And E
=f(P,Q)\And E$.
Then we assume $g(P,Q)=g_1(P,Q)\And g_2(P,Q)$
or $g(P,Q)=g_1(P,Q)\Or g_2(P,Q)$ with two-variable 
polynomials $g_1,g_2$.  If
$g(P,Q)=g_1(P,Q)\And g_2(P,Q)$, 
by the induction hypothesis and distributivity
we have
\beqas
f(P,Q)\And E
&=&
g(P,Q)^\perp\And E\\
&=&
(g_1(P,Q)^\perp\Or g_2(P,Q)^\perp)\And E\\
&=&
(g_1(P,Q)^\perp\And E)
\Or 
(g_2(P,Q)^\perp\And E)\\
&=&
(g_1(P\And E,Q\And E)^\perp\And E)
\Or 
(g_2(P\And E,Q\And E)^\perp\And E)\\
&=&
(g_1(P\And E,Q\And E)^\perp
\Or 
g_2(P\And E,Q\And E)^\perp)\And E)\\
&=&
(g_1(P\And E,Q\And E)
\And
g_2(P\And E,Q\And E))^\perp\And E\\
&=&
g(P\And E,Q\And E)^\perp\And E\\
&=&
f(P\And E,Q\And E)\And E.
\eeqas
Thus, the assertion follows if 
$g(P,Q)=g_1(P,Q)\And g_2(P,Q)$,
and similarly the assertion follows if 
$g(P,Q)=g_1(P,Q)\Or g_2(P,Q)$.
Thus, the assertion generally follows by induction on the complexity of
the polynomial $f$.
\end{proof}

Let $\cL=\{P,Q\}^{!!}$.
Then $[0,\com(P,Q)]$ is 
a complete Boolean algebra
with relative orthocomplement $X^c=X\p\And \com(P,Q)$.
From Proposition \ref{th:absoluteness_commutator},
any $X\in\cL$
is uniquely decomposed as $X=X_B\Or X_N$ with the
condition that $X_B\le\com(P,Q)$ and $X_N\le\com(P,Q)\p$.
Since
 $P^{\al}\And Q^{\be}\le\com(P,Q)$
and $\com(P,Q)\p\le P^{\al}\Or Q^{\be}$, where $\al,\be\in\{\id,\perp\}$,
we have
\beqas
(P^{\al})_B\And (Q^{\be})_B&=&(P^{\al}\And Q^{\be})_B=P^{\al}\And Q^{\be},\\ 
(P^{\al})_N\And (Q^{\be})_N&=&(P^{\al}\And Q^{\be})_N=0,\\
(P^{\al})_B\Or (Q^{\be})_B&=&(P^{\al}\Or Q^{\be})_B=\Sup_{\al':\al'\ne\al;
\be':\be'\ne\be} (P^{\al'}\And Q^{\be'}),\\ 
(P^{\al})_N\Or (Q^{\be})_N&=&(P^{\al}\Or Q^{\be})_N=\com(P,Q)\p.
\eeqas

\begin{prop}\label{th:implications}
Let $\Then$ be a binary operation satisfying  (I1) and (I2).
Then the following conditions are equivalent.
\benum
\item[{\rm (i)}] $\Then$ is a generalized implication, i.e., it satisfies (LB).
\item[{\rm (ii)}] $(P\Then Q)_B=P\Then_0 Q$ for all $P,Q\in\cQ$.
\item[{\rm (iii)}] $(P\Then Q) \Or\com(P,Q)\p = P\Then_5Q$ for all
$P,Q\in\cQ$.
\item[{\rm (iv)}] $P\Then_0Q \le P\Then Q\le P\Then_5 Q$ for all  $P,Q\in\cQ$.
\eenum
\end{prop}
\begin{proof}
Suppose (LB) is satisfied.  Let $P,Q\in\cQ$.
Since $P_B\commutes Q_B$, we have $P_B\Then Q_B
=P_B\p\Or Q_B$ and $(P_B\p\Or Q_B)\And \com(P,Q)=(P\p\Or Q)\And \com(P,Q)=
P\Then_0 Q$.
Thus, from (I2) we have
\beqas
(P\Then Q) \And\com(P,Q)=(P_B\Then Q_B)\And \com(P,Q)
 =P\Then_0 Q,
\eeqas
and hence (i)$\THEN$(ii) follows. 
Suppose (ii) holds.
We have $P\Then_0Q\le P\Then Q$.
By taking the join with $\com(P,Q)^\perp$ in the both sides of relation (ii), we have
$P\Then Q \Or \com(P,Q)^\perp= P\Then_0 Q\Or \com(P,Q)^\perp$.
Since $ P\Then_0 Q\Or \com(P,Q)^\perp=P\Then_5 Q$ by calculation,
we obtain (iii), and the implication (ii)$\THEN$(iii) follows.
Suppose (iii) holds.  
Then $P\Then Q\le P\Then_5 Q$.  By taking the meet with
 $\com(P,Q)$ in the both sides of (iii), we have
$P\Then Q \And \com(P,Q)= P\Then_5 Q\And \com(P,Q)=P\Then_0 Q$,
and hence $P\Then_0 Q\le P\Then Q$.
Thus, the implication (iii)$\THEN$(iv) follows.  
Suppose (iv) holds.  If $P\commutes Q$, we have
$P\Then_0Q =P\Then_5 Q=P\p\Or Q$, so that $P\Then Q=P\p\Or Q$.
Thus, the implication (iv)$\THEN$(i) follows, and the
proof is completed.
\end{proof}

Polynomially definable generalized implications are characterized as follows.

\begin{thm}\label{th:poly}
There are only six polynomially
definable generalized implications, namely, the
six binary operations $\Then_j$ for $j=0,\ldots,5$. 
In particular, they satisfy the following relations for any $P,Q\in \cQ$.
\benum
\item[{\rm (i)}] 
$P\Then_1 Q=(P\Then_0 Q)\Or (P\And \com(P,Q)\p)$.
\item[{\rm (ii)}] 
$P\Then_2 Q=(P\Then_0 Q)\Or (Q\And \com(P,Q)\p)$.
\item[{\rm (iii)}] 
$P\Then_3 Q=(P\Then_0 Q)\Or (P\p\And \com(P,Q)\p)$.
\item[{\rm (iv)}] 
$P\Then_4 Q=(P\Then_0 Q)\Or (Q\p\And \com(P,Q)\p)$.
\item[{\rm (v)}] 
$P\Then_5 Q=(P\Then_0 Q)\Or \com(P,Q)\p$.
\eenum
\end{thm}
\begin{proof}
From Proposition \ref{th:restriction_property} and Kotas's result 
mentioned above \citep{Kot67}, it follows easily that there are only six
polynomially  definable generalized implications, namely, the 
six binary operations $\Then_j$ for $j=0,\ldots,5$. 
From Proposition \ref{th:implications}, we have $(P\Then _j Q)_B=
P\Then_0 Q$ for all $j=0,\ldots,5$.  
Relations (i)--(v) can be easily obtained by the relation
$(P\Then _j Q)_N=(P\Then _j Q)\And \com(P,Q)\p$ for all $j=0,\ldots,5$.
\end{proof}

\begin{thm}\label{th:implication}
Let $\Then$ be a generalized implication on a logic $\cQ$
and let $P,P_1, P_2, P_{1,\al}, P_{2,\al}, Q\in \cQ$.
Then the following statements hold.
\benum
\item[{\rm (i)}]  $P\Then Q=1$ if $P\le Q$.
\item[{\rm (ii)}] $(\Inf_{\al} P_{1,\al}\Then P_{2,\al})\And Q=
(\Inf_{\al} (P_{1,\al}\And Q)\Then(P_{2,\al}\And Q))\And Q$ if 
$P_{1,\al}, P_{2,\al}\commutes Q$.
\eenum
\end{thm}
\begin{proof}
If $P\le Q$, then $P\commutes Q$ and $P\Then Q=P^{\perp}\Or Q=1$,
so that statement (i) follows.
Statement (ii) follows from the definition of generalized implications and 
Proposition  \ref{th:logic}.
\end{proof}

Generalized implications satisfying (MP) are characterized as follows.

\begin{prop}\label{th:quantum_implication}
Let $\Then$ be a generalized implication on a logic $\cQ$.
Then the following conditions are equivalent.
\benum
\item[{\rm (i)}] $\Then$ satisfies (MP).
 \item[{\rm (ii)}] $P\And (P\Then Q)_N=0$ for all $P,Q\in\cQ$.
\eenum

\end{prop}
\begin{proof}
Suppose that (MP) holds.
Then  $P\And(P\Then Q)\le P\And Q$ and hence
\[
P\And(P\Then Q)_N=P\And(P\Then Q)\And\com(P,Q)\p
\le P\And Q\And \com(P,Q)\p=0.
\]
Thus, (ii) holds.
Conversely, suppose that a generalized implication $\Then$ satisfies (ii).
Since $P\Then Q\in \{P,Q\}^{!!}$, from Proposition \ref{th:implications} (ii) we have
\[
P\And(P\Then Q) =(P_B\And(P\Then Q)_B)\Or (P_N\And(P\Then Q)_N)
=P_B\And (P\Then_0 Q)=P\And Q\le Q.
\] 
Thus, (MP) holds, and the proof is completed.
\end{proof}

The following characterization of polynomially definable generalized
implications satisfying (MP) was given by 
\cite{Har81}.

\begin{cor}\label{th:MP}
The only polynomially definable generalized implications satisfying (MP) are
only four binary operations $\Then_j$  for $j=0,2,\ldots,4$.
\end{cor}
\begin{proof}
We have
\beqas
P\And (P\Then_0 Q)_N&=&0,\\
P\And (P\Then_1 Q)_N&=&P\And P_N=P_N,\\
P\And (P\Then_2 Q)_N&=&P\And Q_N=(P\And Q)_N=0,\\
P\And (P\Then_3 Q)_N&=&P\And P\p_N=(P\And P\p)_N=0,\\
P\And (P\Then_4 Q)_N&=&P\And Q\p_N=(P\And Q\p)_N=0,\\
P\And (P\Then_5 Q)_N&=&P\And \com(P,Q)\p=P_N,\\
\eeqas
and the assertion follows from Proposition \ref{th:quantum_implication}.
\end{proof}

The above four implications are mutually characterized 
as follows.

\begin{prop}\label{th:implications2}
 Let $\cQ$  be a logic.
For any $P,Q\in\cQ$, we have the following.
\benum
\item[{\rm (i)}] $X\le P\Then_3 Q$ if and only if 
$P\And (P^{\perp}\Or X)\le Q$.
\item[{\rm (ii)}] $P\Then_3 Q=\max\{X\in\{P\}^{!}\mid
P\And X\le Q\And X\}$.
\item[{\rm (iii)}] $P\Then_2 Q=Q^{\perp}\Then_3 P^{\perp}$.
\item[{\rm (iv)}] $P\Then_2 Q=\max\{X\in\{Q\}^{!}\mid
Q^{\perp}\And X\le P^{\perp}\And X\}$.
\item[{\rm (v)}] $P\Then_0 Q=(P\Then_3 Q)\And
(P\Then_2 Q)$.
\item[{\rm (vi)}] $P\Then_0 Q=\max\{X\in\{P,Q\}^{!}\mid
P\And X\le Q\And X\}$.
\eenum
\end{prop}
\begin{proof}
For the proof of (i), see for example \citep{HMP75}.
Since $P^{\perp}\le (P\Then_3 Q)$,  we have
$(P\Then_3 Q) \commutes P$, and from (MP) we have
$P\Then_3 Q\in\{X\in\{P\}^{!}\mid
P\And X\le Q\}$.
If $X\commutes P$ and $P\And X\le Q$, we have
$$
X=(X\And P)\Or(X\And P^{\perp})\le (P\And Q)\Or P^{\perp}
=P\Then_3 Q. 
$$
Therefore, relation (ii) follows.
Relations (iii) and (iv) are obvious.  
For the proof of (v), see for example \cite[p.~246]{Kal83}.
Since  
$P\And Q, P^{\perp}\And Q, P^{\perp}\And
Q^{\perp}\in \{P,Q\}^{!}$, we have  $P\Then_0 Q\in\{P,Q\}^{!}$.
From (ii), we have 
$P\And (P\Then_0 Q)\le P\And (P\Then_3 Q)\le Q$,
so that $P\Then_0 Q\in\{X\in\{P,Q\}^{!}\mid
P\And X\le Q\}$.
Let $X\in\{P,Q\}^{!}$ and $P\And X\le Q$.
By De Morgan's law, $Q^{\perp}\le P^{\perp}\Or X^{\perp}$.
Since $P\commutes X$, we have
$$
Q^{\perp}\And X\le (P^{\perp}\Or X^{\perp})\And X
=X\And P^{\perp}\le P^{\perp}.
$$  
Thus, by (iv) we have
$X\le P\Then_2 Q$.  We have also $X\le P\Then_3 Q$
from (ii), so that we have $X\le P\Then_0 Q$.
Thus, relation (vi) follows.
\end{proof}

\begin{thm}[Deduction Theorem]\label{th:deduction_theorem}
Let $\Then$ be a generalized implication on a logic $\cQ$.
Then the following statements hold.
\benum
\item[{\rm (i)}] For any $X\in\{P,Q\}^{!}$, if $P\And X\le Q$, then $X\le P\Then
Q$.
\item[{\rm (ii)}] For any $X\in\{P,Q\}^{!}$, we have $\com(P,Q)\And P\And X\le Q$
if and only if $\com(P,Q)\And X\le P \Then Q$.
\item[{\rm (iii)}] $\com(P,Q)\And P\And (P\Then Q)\le Q$.
\eenum
\end{thm}
\begin{proof}
From Proposition \ref{th:implications2} (vi), for any $X\in\{P,Q\}^{!}$, we
have 
$P\And  X\le Q\And  X$ if and only if $X\le P \Then_0 Q$.
It is easy to see that $P \And X \le Q\And X$ if and only if $P \And X \le
Q$. 
Thus, we have $P \And  X \le Q$ if and only if $X \le P \Then_0 Q$, and
assertion (i) follows from
$P\Then_0 Q\le P \Then Q$.
By substituting $X$ by $\com(P,Q) \And  X$, we have $\com(P,Q) \And  P \And
X \le Q$ if
and only if $\com(P,Q) \And  X \le P \Then_0 Q$. Then it is easy to see
that 
$\com(P,Q) \And  X \le P \Then Q$,
since $\com(P,Q) \And  P \Then Q = P \Then_0 Q$. Thus, assertion (ii)
follows. 
Assertion (iii) follows from (ii)
with $X = \com(P,Q) \And  (P \Then Q) = P \Then_0 Q\in\{P,Q\}$.
\end{proof}

Associated with a generalized implication $\Then$ 
we define the {\em logical equivalence} 
by  
$P\Iff Q=(P\Then Q)\And(Q\Then  P)$.
A generalized implication $\Then$ is said to satisfy (LE) if 
 $
P\Iff Q=(P\And Q)\Or(P\p\And Q\p)
$
for all $P,Q\in\cQ$.

\begin{prop}\label{th:LE}
 Let $\Then$ be a generalized implication on a logic $\cQ$. 
Then the following conditions are equivalent.
\benum
\item[{\rm (i)}] $(LE)$ holds.
\item[{\rm (ii)}] $P\Iff Q=\max\{X\in\{P,Q\}^{!}\mid
P\And X= Q\And X\}$.
\item[{\rm (iii)}] $P\Iff Q\le \com(P,Q)$ for all $P,Q\in\cQ$.

In this case, we have
\item[{\rm (iv)}] $P\And (P\Iff Q)\le Q$ for all $P,Q\in\cQ$.
\item[{\rm (v)}] $(P\Iff Q)\And (Q\Iff R)\le P\Iff R$ for all $P,Q,R\in\cQ$.
\eenum
\end{prop}

\begin{proof}
{\em (i) $\THEN$ (ii)}.
Suppose $P\Iff Q=(P\And Q)\Or(P^{\perp}\And Q^{\perp})$.
It is easy to see that $\P\Iff Q\in \{X\in\{P,Q\}^{!}\mid P\And X=Q\And X\}$.
Let $X\in \{P,Q\}^{!}$ be such that $P\And X=Q\And X$.
Then  $X\And P=X\And P\And Q$.  
From $P\And X= Q\And X$, we have 
$P^{\perp}\Or X^{\perp}= Q^{\perp}\Or X^{\perp}$, and hence
$$
X\And P^{\perp}=X\And (P^{\perp}\Or X^{\perp})=
X\And (Q^{\perp}\Or X^{\perp})=X\And Q^{\perp}.
$$
Thus, we have 
$X\And P^{\perp}=X\And P^{\perp}\And
Q^{\perp}$, and hence
$
X=(X\And P)\Or(X\And P^{\perp})=X\And (P\Iff Q).
$
This concludes $X\le(P\Iff Q)$ and  relation (ii) follows from relation (i).

{\em (ii) $\THEN$ (iii).}
Suppose $P\Iff Q=\max\{X\in\{P,Q\}^{!}\mid P\And X= Q\And X\}$.
Then $P\And (P\Iff Q)=Q\And (P\Iff Q)$ and hence 
$P\And (P\Iff Q)\commutes Q\And (P\Iff Q)$.
Thus, $P\Iff Q$ is a subcommutator of $\{P,Q\}$,
and hence $P\Iff Q\le \com(P,Q)$. 

{\em (iii) $\THEN$ (i).}
Suppose $P\Iff Q \le \com(P,Q)$.  Then  
$P\Iff Q = P\Iff Q\And \com(P,Q) 
=(P\Then Q)\And \com(P,Q) \And (Q\Then P)\And \com(P,Q)
= P\Then_0 Q \And Q\Then_0 P=(P\And Q)\Or(P^{\perp}\And Q^{\perp})$.

{\em Proof of (iv)}.
From (ii), we have $P\And (P\Iff Q)=Q\And (P\Iff Q)\le Q$,
and the assertion follows.

{\em Proof of (v)}.
Let $P,Q,R\in \cQ$.  Let $E=P\Iff Q$ and $F=Q\Iff R$.
From (ii) we have $P\And E=Q\And E$ and $Q\And F=R\And F$,
so that $P\And  E\And F=R\And E\And F$.
From (ii) we have $Q\commutes E,F$, so that $Q\commutes E\And F$.
Since $E\commutes E\And F$, we have $Q\And E\commutes E\And F$.
Since $P\And E=Q\And E$, we have $P\And E\commutes E\And F$.
It is obvious that $P\And E\p\commutes E\And F$.  Since
$P\commutes E$, we have $P\commutes E\And F$.  Similarly, we have
$R\commutes E\And F$.  Thus, from (ii) we have $E\And F\le
P\Iff R$, and relation (v) is obtained.
\end{proof}

The following characterization of polynomially definable generalized
implications satisfying (LE) was given by 
\cite{Har81}.

\begin{cor}\label{th:equivalence}
The only polynomially definable generalized implications satisfying (LE) 
are the five binary operations $\Then_j$ for $j=0,\ldots,4$.
\end{cor}
\begin{proof}
From $(P\Iff_j Q)_N=(P\Then_j Q)_N\And (Q\Then_j P)_N$,
we have
\beqas
(P\Iff_0 Q)_N&=&0,\\
(P\Iff_1 Q)_N&=&P_N\And Q_N=(P\And Q)_N=0,\\
(P\Iff_2 Q)_N&=&Q_N\And P_N=(Q\And P)_N=0,\\
(P\Iff_3 Q)_N&=&P\p_N\And Q\p_N=(P\p\And Q\p)_N=0,\\
(P\Iff_4 Q)_N&=&Q\p_N\And P\p_N=(Q\p\And P\p)_N=0,\\
(P\Iff_5 Q)_N&=&\com(P,Q)\p.
\eeqas
From Proposition \ref{th:LE} (iii), the generalized 
implication $\Then_j$ satisfies (LE) if and only if
$(P\Iff_j Q)_N=0$, and the assertion follows.
\end{proof}

\xsection{Non-polynomial implications in quantum logic.}
\label{se:NPIIQL}
 
In the preceding section, we introduced the notion of generalized implications.
In this section, we shall show that there are continuously many generalized 
implications on the projection lattices of von Neumann algebras
definable by the general structure of von Neumann algebras
but not definable as an ortholattice polynomial.

\cite{BK73} determined 
the structure of the subalgebra $\Ga_0\{P,Q\}$ generated by $P,Q\in\cQ$ 
to be isomorphic to the direct product of a Boolean
algebra and MO2=$\{0,a,a^{\perp},b,b^{\perp},1\}$, 
the Chinese lantern \cite[p.~16, p.~27]{Kal83}.
In this case, $\Ga_0\{P,Q\}$ is a complete subalgebra so that 
$\Ga_0\{P,Q\}=\Ga\{P,Q\}$, and $[0,\com(P,Q)]_{\Ga\{P,Q\}}$ is a 
Boolean algebra and $[0,\com(P,Q)^{\perp}]_{\Ga\{P,Q\}}$ is 
isomorphic to MO2.
However, the structure of the sublogic $\{P,Q\}^{!!}$ generated by  $P,Q\in\cQ$ 
is more involved.  
For the projection lattice $\cQ=\cP(\cM)$ of a von Neumann algebra $\cM$,
the sublogic $\{P,Q\}^{!!}$ is the projection lattice of the von Neumann algebra 
$\{P,Q\}''$ generated by $P,Q\in\cQ$ \citep{07TPQ}.  
For example, let $P,Q\in\cQ(\cH)=\cP({\rm B}(\cH))$ 
be rank one projections on a Hilbert space $\cH$.
Then  $\com(P,Q)=1$ or $\com(P,Q)=0$.
If $P=Q$ or $P\perp Q$, then  $\com(P,Q)=1$ and $\{P,Q\}^{!!}=\Ga\{P,Q\}$ is a
complete Boolean subalgebra of
$\cQ$. Otherwise, $\com(P,Q)=0$ and $\{P,Q\}^{!!}$ is isomorphic to 
$\cQ(\C^2)=\cP({\rm B}(\C^2))$, 
but $\Ga\{P,Q\}$ is a 6-element subalgebra of $\{P,Q\}^{!!}$ isomorphic to MO2.
Thus, $\{P,Q\}^{!!}$ is much larger than $\Ga\{P,Q\}$.
This is an example in which a complete subalgebra is not a sublogic.

Define a binary operation $\circ_{\theta}$ on the projection lattice $\cQ=\cP(\cM)$ 
of a von Neumann algebra $\cM$ by 
\beqas
P\circ_{\theta}Q=e^{i\theta P}Q e^{-i\theta P}
\eeqas
for all $P,Q\in\cQ$.  
If $P\commutes Q$, then we have $P\circ_{\theta}Q=Q$.
We have 
\beqas
P\circ_{\theta}Q=Q+(e^{i\theta}-1)PQ+(e^{-i\theta}-1)QP+2(1-\cos\theta)PQP
\eeqas
for all $P,Q\in\cQ$.  This was first introduced by 
\cite{Ta81} for 
$\cM={\rm B}(\cH)$.  Then the binary operation $f(P,Q)=P\circ_{\theta} Q$
satisfies conditions (i) and (ii)  in Proposition \ref{th:restriction_property}.
However, it is not in general definable as a lattice polynomial, since $f(P,Q)$ is
not generally in $\Ga\{P,Q\}$ as shown in the proof of Proposition \ref{th:non-poly}
below.

Now, for $j=0,\ldots,5$, for a real parameter $\theta\in [0,2\pi)$, and for $i=0,1$,
we define binary operations $\Then_{j,\theta,i}$
on
$\cQ=\cP(\cM)$ by 
\beqas
P\Then_{j,\theta,0}Q&=&P\Then_{j}(P\circ_{\theta} Q)\\
P\Then_{j,\theta,1}Q&=&(Q\circ_{\theta}P)\Then_{j} Q
\eeqas
for all $P,Q\in\cQ$.  Obviously, $\Then_{j,0,i} =\Then_{j}$ for $j=0,\ldots,5$
and $i=0,1$.

\begin{prop}
For any von Neumann algebra $\cM$,
the binary operations $\Then_{j,\theta,i}$ on $\cQ=\cP(\cM)$ for
$j=0,\ldots,5$, $\theta\in[0,2\pi)$, and $i=0,1$ are generalized implications.
In particular, they satisfy the following relations for any $P,Q\in \cQ$
and $\theta\in[0,2\pi)$.

\benum
\item[\rm (i)] $P\Then_{0,\theta,0} Q=P\Then_0 Q$.
\item[\rm (ii)] $P\Then_{1,\theta,0} Q=P\Then_1 Q$.
\item[\rm (iii)] $P\Then_{2,\theta,0} Q
=(P\Then_0 Q)\Or (P\circ_{\theta}Q\And \com(P,Q)\p)$.
\item[\rm (iv)] $P\Then_{3,\theta,0} Q=P\Then_3 Q$.
\item[\rm (v)] $P\Then_{4,\theta,0} Q=(P\Then_0 Q)\Or (P\circ_{\theta}Q\p\And \com(P,Q)\p)$.
\item[\rm (vi)] $P\Then_{5,\theta,0} Q=P\Then_5 Q$.
\item[\rm (vii)] $P\Then_{0,\theta,1} Q=P\Then_0 Q$.
\item[\rm (viii)] $P\Then_{1,\theta,1} Q=(P\Then_0 Q)\Or (Q\circ_{\theta}P\And \com(P,Q)\p)$.
\item[\rm (ix)] $P\Then_{2,\theta,1} Q=P\Then_2 Q$.
\item[\rm (x)] $P\Then_{3,\theta,1} Q=(P\Then_0 Q)\Or (Q\circ_{\theta}P\p\And \com(P,Q)\p)$.
\item[\rm (xi)] $P\Then_{4,\theta,1}Q=P\Then_4 Q$.
\item[\rm (xii)] $P\Then_{5,\theta,1} Q=P\Then_5 Q$.
\eenum
\end{prop}
\begin{proof}
We have
$$
(P\Then_{j,\theta,0}Q)_B=P\circ_{\theta} (P\Then_{j}Q)_B
=P\circ_{\theta} (P\Then_{0}Q)
=P\Then_{0}Q
$$
and
$$
(P\Then_{j,\theta,1}Q)_B=Q\circ_{\theta} (P\Then_{j}Q)_B
=Q\circ_{\theta} (P\Then_{0}Q)
=P\Then_{0}Q
$$
for all $j=0,\ldots,5$.  
It follows from Proposition \ref{th:implications} (ii)
that $\Then_{j,\theta,i}$ is a generalized implication
for all $j=0,\ldots,5$, $\theta\in[0,2\pi)$, and $i=0,1$.
We have 
$$
(P\Then_{j,\theta,0}Q)_N=P\circ_{\theta} (P\Then_{j}Q)_N,
$$
and hence
\beqas
(P\Then_{0,\theta,0}Q)_N&=&0,\\
(P\Then_{1,\theta,0}Q)_N&=&P\circ_{\theta} (P\And\com(P,Q)^{\perp})
=P\And\com(P,Q)^{\perp}=(P\Then_{1}Q)_N,\\
(P\Then_{2,\theta,0}Q)_N&=&P\circ_{\theta} (Q\And\com(P,Q)^{\perp})
=P\circ_{\theta} Q\And\com(P,Q)^{\perp},\\
(P\Then_{3,\theta,0}Q)_N&=&P\circ_{\theta} (P^{\perp}\And\com(P,Q)^{\perp})
=P^{\perp}\And\com(P,Q)^{\perp}=(P\Then_{3}Q)_N,\\
(P\Then_{4,\theta,0}Q)_N&=&P\circ_{\theta} (Q^{\perp}\And\com(P,Q)^{\perp})
=P\circ_{\theta} Q^{\perp}\And\com(P,Q)^{\perp},\\
(P\Then_{5,\theta,0}Q)_N&=&P\circ_{\theta}\com(P,Q)^{\perp}
=\com(P,Q)^{\perp}.
\eeqas
Thus, we obtain relations (i)--(vi).  The rest of the assertions follow
similarly.
\end{proof}

In what follows, for any two vectors $\xi,\et$ in a Hilbert space $\cH$
the operator $\ket{\xi}\bra{\et}$ is defined by $\ket{\xi}\bra{\et}\ps
=\bracket{\et|\ps}\xi$ for all $\ps\in\cH$,
where $\bracket{\cdots|\cdots}$ stands for the inner product of $\cH$,
which is assumed to be linear in the second variable. 
If $\xi$ or $\et$ are denoted by $\ket{a}$ or $\ket{b}$, respectively, 
 as is customary in quantum mechanics \citep{Dir58},
the inner product $\bracket{\xi|\et}$ 
is also denoted by $\bracket{a|b}$, $\bracket{a|\et}$, or $\bracket{\xi|b}$,
and the  operator $\ket{\xi}\bra{\et}$ is also denoted by $\ket{a}\bra{b}$,
$\ket{a}\bra{\et}$, or $\ket{\xi}\bra{b}$.

\begin{prop}\label{th:non-poly}
Generalized implications $\Then_{1,\theta,1}$, $\Then_{2,\theta,0}$, 
$\Then_{3,\theta,1}$, and $\Then_{4,\theta,0}$ are definable on
the projection lattice of an arbitrary von Neumann algebra,
but it is not polynomially definable for any $\theta\in(0,2\pi)$.
\end{prop}
\begin{proof}
Let $\cM={\rm B}(\C^2)$ and let $\{\ket{0},\ket{1}\}$ be a complete orthonormal
basis of $\C^2$.  Let $\vp= (1/2)(\ket{0}+\sqrt{3}\ket{1})$.
Let $\theta\in(0,2\pi)$.
Let $P=\ketbra{\vp}$, and $Q=\ketbra{1}$.
Then  $Q\circ_{\theta}P=\ketbra{\vp(\theta)}$
where $\vp(\theta)=(1/2)(\ket{0}+e^{i\theta}\sqrt{3}\ket{1})$.
Since $\bracket{1|\vp}=\sqrt{3}/2$, we have
$\com(P,Q)=0$.  Thus, 
$$
P\Then_{1,\theta,1}Q=Q\circ_{\theta}P=\ketbra{\vp(\theta)}.
$$
Since $\bracket{\vp|\vp(\theta)}= (1+3e^{i\theta})/4$ and 
$\bracket{1|\vp(\theta)}=\sqrt{3}e^{i\theta}/2$,  it follows that
$P\circ_{\theta}Q$ is not an element of $\{0,P,P\p,Q,Q\p,1\}$.
Since the subalgebra $\Ga\{P,Q\}$ generated by $P,Q$ is a Chinese 
lantern $\{0,P,P\p,Q,Q\p,1\}$, we conclude that there is no
ortholattice polynomial $f(P,Q)$ such that $f(P,Q)=P\Then_{1,\theta,1} Q$
holds in any $\cP(\cM)$.
The rest of the assertion can be proved similarly.
\end{proof}

\begin{prop} 
For any von Neumann algebra $\cM$,
the binary operations $\Then_{j,\theta,i}$ on $\cQ=\cP(\cM)$ with
$j=0,2,\ldots,4$, $\theta\in[0,2\pi)$, and $i=0,1$ but
$(j,i)\not=(3,1)$ satisfy {\rm (MP)}.
\end{prop}
\begin{proof}
For $(j,i)=(0,0), (0,1), (2,1), (3,0), (4,1)$, we have $\Then_{j,\theta,i}=
\Then_j$, and hence the assertion follows from Proposition \ref{th:MP}.
For $(j,i)=(4,0)$, we have
$$
P\And (P\Then_{4,\theta,0} Q)_N=P\And (P\circ_{\theta} Q\p)_N=
P\circ_{\theta}(P\And Q\p)_N=0,
$$
and hence $\Then_{4,\theta,0}$ satisfies (MP) by Proposition 
\ref{th:quantum_implication}.
For $(j,i)=(2,0)$ the assertion can be verified analogously.
\end{proof}

\xsection{Universe of quantum sets.}
\label{se:UOQS}

Let $\cQ$ be an arbitrary complete orthomodular
lattice.
We denote by $\V$ the universe of 
the Zermelo-Fraenkel set theory
with the axiom of choice (ZFC).
Throughout this paper, 
we fix the language  $\LL_{\in}$
for first-order theory with equality augmented by
a binary relation symbol
$\in$, bounded quantifier symbols $\forall x\in y$,
$\exists x \in y$, and no constant symbols.
For any class $U$, 
the language $\LL_{\in}(U)$ is the one
obtained by adding a name for each element of $U$.
We consider $\Not$, $\And$, $\Then$, $\forall x\in y$,
$\exists x \in y$, and $(\forall x)$  as primitive symbols,
while $\Or$, $\Iff$, and  $(\exists x)$  as derived symbols in the obvious ways. 
For convenience, 
we use the same symbol for an element of $U$ and
its name in $\LL_{\in}(U)$ as well as for the membership
relation and the symbol $\in$.

To each sentence $\vp$ of $\LL_{\in}(U)$, the satisfaction
relation
$\bracket{U,\in} \models \vp$ is defined by the following recursive rules:
\benum
\item[\rm (i)] $ \bracket{U,\in} \models u\in v
\IFF u\in v.$
\item[\rm (ii)] $ \bracket{U,\in} \models u = v
\IFF u = v.$ 
\item[\rm (iii)] $ \bracket{U,\in} \models \Not \vp \IFF  \bracket{U,\in}
\models
\vp 
\mbox{ does not hold}$. 
\item[\rm (iv)] $ \bracket{U,\in} \models \vp_1 \And \vp_2 
\IFF \bracket{U,\in}
\models \vp_1 
\mbox{ and } \bracket{U,\in} \models \vp_2$.
\item[\rm (v)] $ \bracket{U,\in} \models \vp_1 \Then \vp_2 
\IFF \bracket{U,\in}
\models \vp_1 \mb{ does not hold or }\bracket{U,\in}\models \vp_2.$ 
\item[\rm (vi)] $ \bracket{U,\in} \models  (\forall x\in u)\,\vp(x) \IFF
\bracket{U,\in} \models \vp(u') \mbox{ for all } u' \in u$ .
\item[\rm (v)] $ \bracket{U,\in} \models  (\exists x\in u)\,\vp(x) \IFF
\mb{ there exists $u'\in u$ such that
$\bracket{U,\in} \models \vp(u')$.}$
\item[\rm (vi)] $ \bracket{U,\in} \models  (\forall x)\,\vp(x) \IFF
\bracket{U,\in} \models \vp(u) \mbox{ for all } u \in U$ .
\eenum
Our assumption that $\V$ satisfies ZFC
means that if $\vp(x_1,\ldots,x_n)$ is provable in 
ZFC,  i.e., 
$\mb{ZFC}\vdash \vp(x_1,\ldots,x_n)$, then 
$\bracket{\V,\in}\models \vp(u_1,\ldots,u_n)$ for 
any formula $\vp(x_1,\ldots,x_n)$ of $\LL_{\in}$ and
all $u_1,\ldots,u_n\in \V$. 

Let $\cQ$ be a logic.
By transfinite recursion, 
we define $\V_{\al}^{(\cQ)}$  for each ordinal ${\al}$ by
$$
\V_{\al}^{(\cQ)} = \{u|\ u:\dom(u)\to \cQ \mbox{ and }
(\exists \be<\al)
\dom(u) \subseteq V_{\be}^{(\cQ)}\}.
$$
Thus, each element of $\V_{\al}^{(\cQ)}$ is a $\cQ$-valued 
function defined on a subset of  $V_{\be}^{(\cQ)}$ for some $\be<\al$.
We have $\V_{0}^{(\cQ)}=\emptyset$, $\V_{1}^{(\cQ)}=\{\emptyset\}$,
$\V_{2}^{(\cQ)}=\{\emptyset\}\cup\{\bracket{\emptyset,P}\mid P\in\cQ\}$,
and so on.
The {\em $\cQ$-valued universe} $\VQ$ is defined
by 
$$
  \VQ= \bigcup _{{\al}{\in}{\rm On} }V_{{\al}}^{(\cQ)},
$$
where $\mbox{On}$ is the class of ordinals. 
It is easy to see that if $\cL$ is a sublogic of $\cQ$
then 
$\VL_{\al}\subseteq\VQ_{\al}$
for all $\al$.
For every $u\in\VQ$, the {\em rank} of $u$, denoted by
$\rank(u)$,  is defined as the least $\al$ such that
$u\in \VQ_{\al+1}$.
It is easy to see that if $u\in\dom(v)$ then 
$\rank(u)<\rank(v)$

For $u\in\VQ$, we define the {\em support} 
of $u$, denoted by $S(u)$, by transfinite recursion on the 
rank of $u$ with the relation
$$
S(u)=\bigcup_{x\in\dom(u)}S(x)\cup\{u(x)\mid x\in\dom(u)\}.
$$
For $\cU\subseteq\VQ$ we write 
$S(\cU)=\bigcup_{u\in\cU}S(u)$ and
for $u_1,\ldots,u_n\in\VQ$ we write 
$S(u_1,\ldots,u_n)=S(\{u_1,\ldots,u_n\})$
and $S(\vec{u})=S(u_1,\ldots,u_n)$ if $\vec{u}=(u_1,\ldots,u_n)$.
Then we obtain the following characterization of
subuniverses of $\VQ$.

\begin{prop}\label{th:sublogic}
Let $\cL$ be a sublogic of $\cQ$ and $\al$ an
ordinal. For any $u\in \VQ$, we have
$u\in\VL_{\al}$  if and only if
$u\in \VQ_{\al}$ and $S(u)\in\cL$.  
In particular, $u\in\VL$ if and only if
$u\in\VQ$ and $S(u)\in\cL$. 
Moreover, if $u\in\VL$, then
$\rank(u)$ defined in $\VL$ and the one 
defined in $\VQ$ are the same. 
\end{prop}
\begin{proof}  Immediate from transfinite induction on
$\al$.
\end{proof}

Let $\Then$ be an arbitrary generalized implication on $\cQ$ and define 
$\Iff$ by $P\Iff Q=(P\Then Q)\And(Q\And P)$ for all $P,Q\in\cQ$;
the same symbols will be used for the corresponding logical connectives for 
implication and logical equivalence.  
To each sentence $\vp$ of $\LL_{\in}(\VQ)$ 
we assign the
$\cQ$-valued truth value $ \val{\vp}$,
called the {\em $(\cQ,\Then)$-valued interpretation} of $\vp$, 
by the following
recursive rules:
\benum
\item[\rm (i)] $\val{u = v}
= \inf_{u' \in  \dom(u)}(u(u') \Then
\val{u'  \in v})
\And \inf_{v' \in   \dom(v)}(v(v') 
\Then\val{v'  \in u})$.
\item[\rm (ii)] $ \val{u \in v} 
= \sup_{v' \in \dom(v)} (v(v') \And \val{v'=u})$.
\item[\rm (iii)] $ \val{\Not\vp} = \val{\vp}^{\perp}$.
\item[\rm (iv)] $ \val{\vp_1\And\vp_2} 
= \val{\vp_{1}} \And \val{\vp_{2}}$.
\item[\rm (v)] $ \val{\vp_1\Or\vp_2} 
= \val{\vp_{1}} \Or \val{\vp_{2}}$.
\item[\rm (vi)] $ \val{\vp_1\Then\vp_2} 
= \val{\vp_{1}} \Then\val{\vp_{2}}$.
\item[\rm (vii)] $ \val{\vp_1\Iff\vp_2} 
= \val{\vp_{1}} \Iff \val{\vp_{2}}$.
\item[\rm (viii)] $ \val{(\forall x\in u)\, {\vp}(x)} 
= \Inf_{u'\in \dom(u)}
(u(u') \Then \val{\vp(u')})$.
\item[\rm (ix)] $ \val{(\exists x\in u)\, {\vp}(x)} 
= \Sup_{u'\in \dom(u)}
(u(u') \And \val{\vp(u')})$.
\item[\rm (x)] $ \val{(\forall x)\, {\vp}(x)} 
= \Inf_{u\in \VQ}\val{\vp(u)}$.
\item[\rm (xi)] $ \val{(\exists x)\, {\vp}(x)} 
= \Sup_{u\in \VQ}\val{\vp(u)}$.
\eenum

In the above relations (i) and (ii) can be considered as a definition of
 $\val{u = v}$ and $\val{u \in v}$ by recursion on a well-founded relation
 such that
 \beqas
 \mb{
 $\bracket{u,v}<\bracket{u',v'}$ if and only if either ($u\in\dom(u')$ and $v=v'$) 
 or ($u=u'$ and $v\in\dom(v')$ holds.
 }
\eeqas
See \citep[p.~23]{Bel05} for details and \citep[pp.~121--122]{TZ73} for alternative ways to check that
(i) and (ii) constitute a definition by recursion.
Then relations (iii)--(viii) define  $ \val{\vp}$ for all sentences $\vp$ of $\LL_{\in}(\VQ)$ 
by induction on the complexity of 
$\vp$.

We say that a sentence 
 ${\vp}$ of $ \LL_{\in}(\VQ) $ 
{\em holds} in the
$(\cQ,\Then)$-valued interpretation
 if $ \val{{\vp}} = 1$.
 
De Morgan's laws are satisfied as follows.
\benum
\item[(D1)] $\val{\Not(\vp_{1} \Or \vp_{2})}
=\val{\Not\vp_{1}\And\Not\vp_{2}},\quad 
\val{\Not(\vp_{1} \And \vp_{2})}
=\val{\Not\vp_{1}\Or\Not\vp_{2}}.$
\item[(D2)] $\val{\Not(\exists x)\, {\vp}(x)} 
=\val{(\forall x)\, \Not{\vp}(x)},\quad
 \val{\Not(\forall x)\, {\vp}(x)} 
=\val{(\exists x)\, \Not{\vp}(x)}.$
\eenum

However, it is only in the case where the operation
$\Then$ on $\cQ$ is the maximum
implication $\Then_5$ that 
De Morgan's laws hold for bounded quantifiers:
\benum
\item[(D3)] $\val{\Not(\exists x\in u)\, {\vp}(x)} 
= \val{(\forall x\in u)\,\Not{\vp}(x)},\quad 
\val{\Not(\forall x\in u)\, {\vp}(x)} 
= \val{(\exists x\in u)\,\Not{\vp}(x)}.$
\eenum

According to the theory of Boolean-valued models for 
set theory \citep{Bel05},
for any complete Boolean algebra $\cB$ the Boolean-valued universe
$V^{(\cB)}$  is defined in the same way as $V^{(\cQ)}$ for $\cQ=\cB$.
Since the generalized implication $\Then$ satisfies
$P\Then Q=P^{\perp}\Or Q$ for all $P,Q\in \cB$ by (LB),
it is easy to see that our definition of the truth value $\val{\vp}$ 
coincides with the definition in the theory of 
Boolean-valued models for any sentence $\vp$ in $\LL_{\in}(\VB)$,
if $\vp$ does not contain bounded quantifier $(\forall x\in y)$
or $(\exists x\in y)$.
The next proposition shows that even for bounded quantifiers 
we have no conflict.

\begin{prop}
If $\cQ$ is a Boolean logic, then
for any formula ${\vp(x)} $ of $\LL_{\in}(\VQ)$, 
we have
\beqas
\val{(\forall x\in u)\vp(x)}
&=&
\val{(\forall x)x\in u \Then \vp(x)},\\
\val{(\exists x\in u)\vp(x)}
&=&
\val{(\exists x)x\in u \And \vp(x)}.
\eeqas
\end{prop}
\begin{proof}
According to the theory of Boolean-valued models,
if $\cQ$ is Boolean,  we have
\beqas
\val{(\forall x\in u)\vp(x)}
&=&
\Inf_{u'\in \dom(u)}
(u(u') \Then \val{\vp(u')})
=
\Inf_{u'\in\VQ}(\val{u'\in u}\Then \val{\vp(u')})\\
&=&
\val{(\forall x)x\in u \Then \vp(x)},\\
\val{(\exists x\in u)\vp(x)}
&=&
\Sup_{u'\in \dom(u)}
(u(u') \And \val{\vp(u')})
=
\Sup_{u'\in\VQ}(\val{u'\in u} \And \val{\vp(u')})\\
&=&
\val{(\exists x)x\in u \And \vp(x)}.
\eeqas
\end{proof}

The following theorem is an important consequence of 
the axiom of choice \cite[Lemma 1.27]{Bel05}
\begin{thm}[Boolean Maximum Principle]\label{th:2.3.3}  
If $\cQ$ is a Boolean logic,
for any formula $\vp(x)$ of $\LL_{\in}(\VQ)$,
there exists some $u\in\VQ$  such that 
$$
\val{{\vp}(u)} = \val{({\exists}x)\,{\vp}(x)}.
$$ 
\end{thm}

 The basic theorem on Boolean-valued universes is the 
following \cite[Theorem 1.33]{Bel05}.

\begin{thm}[Boolean Transfer Principle]\label{th:BTP}  
\sloppy
If $\cQ$ is a Boolean logic, then
for any formula $\vp(x_{1},\ldots,x_{n})$ of $\LL_{\in}$ and all
$u_{1},\ldots,u_{n}\in\VB$, if 
$\mb{ZFC}\vdash \vp(x_{1},\ldots,x_{n})$ then
$
\val{\vp(u_{1},\ldots,u_{n})}=1
$.
\end{thm} 

A formula in $\LL_{\in}$ 
is called a {\em
$\De_{0}$-formula}  if it has no unbounded quantifier
$\forall x$ or $\exists x$.
For a sublogic $\cL$ of $\cQ$ and a sentence $\vp$ in $\LL_{\in}(V^{(\cL)})$, 
we denote by $\val{\vp}_{\cL}$ the truth value
of $\vp$ defined through $V^{(\cL)}$.

\begin{thm}[$\De_{0}$-Absoluteness Principle]
\label{th:Absoluteness}
\sloppy  
Let $\cL$ be a sublogic of a logic $\cQ$.
For any $\De_{0}$-sentence 
${\vp}$  of  $\LL_{\in}(V^{(\cL)})$, 
we have
$\val{\vp}_{\cL}=\val{\vp}$.
\end{thm}
\begin{proof}
The proof is analogous to the proof of Theorem 3.2 in \cite{07TPQ}.
\end{proof}

The universe $\V$  can be embedded in
$\VQ$ by the following operation 
$\vee:v\mapsto\check{v}$ defined by
$\check{v} = \{\check{u}|\ u\in v\} 
\times \{1\}$ for each $v\in\V$
recursively on the well-founded relation $\in $.
Then the following theorem,  an immediate consequence of
the $\De_0$-Absoluteness Principle,
shows that the subclass $\{\check{x}\mid x\in V\}\subseteq \VQ$ 
is a submodel of $\VQ$ elementarily equivalent to 
$V$ for $\De_0$-formulas in $\LL_{\in}(V)$.

\begin{thm}[$\De_0$-Elementary Equivalence Principle]
\label{th:2.3.2}
For any $\De_{0}$-formula 
${\vp} (x_{1},{\ldots}, x_{n}) $ 
of $\LL_{\in}$ and $u_{1},{\ldots}, u_{n}\in V$,
we have
$
\vp(u_{1},\ldots,u_{n})
$ holds if and only if 
$
\val{\vp(\check{u}_{1},\ldots,\check{u}_{n})}=1.
$
\end{thm}
\begin{proof}
Analogous to \cite[Theorem 3.3]{07TPQ}.
\end{proof}
 
\begin{prop}\label{th:Equality}
For any $u,v \in\VQ$, the following relations hold.
\benum 
\item[\rm (i)] $\val{u=v}=\val{v=u}$.
\item[\rm (ii)] $\val{u=u}=1$.
\item[\rm (iii)] $u(x)\le \val{x\in u}$ for any $x\in\dom(u)$.
\eenum
\end{prop}
\begin{proof}
Relation (i) is obvious from the symmetry of the definition.
We shall prove relations (ii) and (iii) by transfinite induction on 
the rank of $u$.  The relations trivially hold  if $u$ is of the lowest
rank.  
Let $u\in\VQ$. 
We assume that the relations hold for those with lower rank than $u$. 
Let $x\in\dom(u)$. 
By induction hypothesis we have $\val{x=x}=1$, so that  we have
$$
\val{x\in u}=\Sup_{y\in\dom(u)}(u(y)\And\val{x=y})
\ge u(x)\And\val{x=x}=u(x).
$$
Thus, assertion (iii) holds for $u$.  
Then  $(u(x)\Then\val{x\in u})=1$ for all $x\in\dom(u)$,
and hence $\val{u=u}=1$ follows
from Theorem \ref{th:implication} (i).  
Thus, relations (ii) and (iii) hold
by transfinite induction.
\end{proof}

\cite{Tit99} and \cite{TK03}
constructed the lattice-valued universe $V^{(\cL)}$ for 
any complete lattice $\cL$ in the same way as Boolean-valued
universes.  They developed a lattice-valued set theory 
with implication $\Then_{T}$ and negation $\Not_{T}$
defined as follows: $P\Then_{T} Q=1$ if $P\le Q$,
and $P\Then_{T} Q=0$ otherwise;
$\Not_{T} P=1$ if $P=0$, and 
$\Not_{T} P=0$ otherwise, where $P,Q\in\cL$.
This theory can be applied to complete orthomodular lattices,
but the implication $\Then_{T}$ does not generally satisfy the requirements
for generalized implications, in particular (LB), 
and the negation $\Not_{T}$ is different
from the orthocomplementation.
Although their theory includes the case where $\cL$ is a complete Boolean algebra
$\cB$, the truth value defined in their theory is different from
the one defined in the theory of Boolean-valued models, if $\cB\ne {\bf 2}$,
in contrast to the present theory.

\xsection{Transfer principle in quantum set theory.}
\label{se:ZFC}\label{se:TPIQST}

Throughout this section,
 let $\cQ$ be a logic with a generalized implication $\Then$.
Let $u\in\VQ$ and $p\in\cQ$.
The {\em restriction} $u|_p$ of $u$ to $p$ is defined by
the following transfinite recursion on the rank of $u\in\VQ$:
$$
u|_p=\{\bracket{x|_p,u(x)\And p} \mid x\in\dom(u)\}.
$$
By induction, it is easy to see that 
$(u|_p)|_q=u|_{p\And q}$
for all $u\in\VQ$.

In general,  any mapping $\vp:\cQ\to\cQ$ can be naturally lifted up to 
a mapping $\hvp:\VQ\to\VQ$ by transfinite recursion on the rank of $u\in\VQ$:
$$
\hvp(u)=\{\bracket{\hvp(x),\vp[u(x)]} \mid x\in\dom(u)\}.
$$
The restriction $u\mapsto u|_p$ lifts up the mapping $P\in \cQ\mapsto P\And p\in \cQ$
to a mapping $\VQ\to\VQ$ in this way.

\begin{prop}\label{th:L-restriction}
For any $\cU\subseteq \VQ$ and $p\in\cQ$, 
we have 
$$
S(\{u|_p\mid u\in\cU\})=S(\cU)\And p.
$$
\end{prop}
\begin{proof}
By induction, it is  easy to see 
$
S(u|_p)=S(u)\And p,
$
so the assertion follows easily.
\end{proof}

Let $\cU\subseteq\VQ$.  The {\em logic
generated by $\cU$}, denoted by $L(\cU)$, is  define by 
$$
L(\cU)=S(\cU)^{!!}.
$$
For $u_1,\ldots,u_n\in\VQ$, we write
$L(u_1,\ldots,u_n)=L(\{u_1,\ldots,u_n\})$.

\begin{prop}\label{th:commutativity}
For any $\De_0$-formula $\vp(x_1,\ldots,x_n)$ in
$\LL_{\in}$ and $u_1,\cdots,u_n\in\VQ$,
we have the following.
\benum
\item[\rm (i)] $\val{\vp(u_1,\ldots,u_n)}\in L(u_1,\ldots,u_n)$.
\item[\rm (ii)] If 
$p\in S(u_1,\ldots,u_n)^{!}$, then 
$p\commutes \val{\vp(u_1,\ldots,u_n)}$
and $p\commutes \val{\vp(u_1|_p,\ldots,u_n|_p)}$.
\eenum
\end{prop}
\begin{proof}
Analogous to the proofs of Propositions 4.2 and 4.3 in
\cite{07TPQ}. 
\end{proof}

We define the binary relation $x_1\subseteq x_2$ by
``$x_1\subseteq x_2$''=``$\forall x\in x_1(x\in x_2)$.''
Then by definition, for  any $u,v\in\VQ$ we have
$$
\val{u\subseteq v}=
\Inf_{u'\in\dom(u)}
u(u')\Then \val{u'\in v},
$$
and $\val{u=v}=\val{u\subseteq v}
\And\val{v\subseteq u}$.

\begin{prop}\label{th:restriction-atom}
For any $u,v\in\VQ$ and $p\in S(u,v)^{!}$, we have
the following relations.
\benum
\item[\rm (i)] $\val{u|_p\in v|_p}=\val{u\in v}\And p$.
\item[\rm (ii)] $\val{u|_p\subseteq v|_p}\And p
=\val{u\subseteq v}\And p$.
\item[\rm (iii)] $\val{u|_p= v|_p}\And p =\val{u= v}\And p$
\eenum
\end{prop}
\begin{proof}
We prove these relations by induction on the ranks of 
$u,v$.  If $\rank(u)=\rank(v)=0$, then $\dom(u)=\dom(v)
=\emptyset$, so that the relations trivially hold.
Let $u,v\in\VQ$ and $p\in S(u,v)^{!}$.
To prove (i), suppose $v\in\VQ_{\al}$, $u\in\VQ_{\be}$, $\be<\al$, 
and $p\in S(u,v)^{!}$.
Let $v'\in\dom(v)$. 
Then $p\commutes v(v')$ by the assumption on $p$.
By Proposition \ref{th:commutativity} (ii), we have 
$p\commutes \val{u=v'}$, 
and hence $v(v')\And\val{u=v'}\commutes p$. 
By induction hypothesis, we also have 
$\val{u|_p=v'|_p}\And p=\val{u=v'}\And p$.
Thus,  we  have
$$
\val{u|_p\in v|_p}
=
\Sup_{v'\in\dom(v)}
v|_p(v'|_p)\And\val{u|_p=v'|_p}\\
=
\Sup_{v'\in\dom(v)}
v(v')\And \val{u=v'}\And p.
$$
From Proposition \ref{th:logic} we obtain
$$
\val{u|_p\in v|_p}=
\left(\Sup_{v'\in\dom(v)}
v(v')\And \val{u=v'}\right)\And p.
$$
Thus, we obtain  relation (i) by the definition of $\val{u=v}$. 
To prove (ii),  suppose $u,v\in\VQ_{\al}$ and $p\in S(u,v)^{!}$.
Let $u'\in\dom(u)$.
Then  $\val{u'|_p\in v|_p}=\val{u'\in v}\And p$ by relation (i).
Thus, we have
$$
\val{u|_p\subseteq v|_p}
=
\Inf_{u'\in\dom(u)}
(u(u')\And p)\Then(\val{u'\in v}\And p).
$$
We  have $p\commutes u(u')$ by
assumption on $p$, and $p\commutes\val{u'\in v}$
by Proposition \ref{th:commutativity} (ii).
By property (I2) of generalized implications, we have
$$
p\And [(u(u')\And p)\Then(\val{u'\in v}\And p)]=
p\And (u(u')\Then\val{u'\in v}).
$$
Thus, by Proposition \ref{th:implication} (ii) we have
\beqas
p\And\val{u|_p\subseteq v|_p}
&=&
p\And\Inf_{u'\in\dom(u)}
(u(u')\And p)\Then(\val{u'\in v}\And p)\\
&=&
p\And\Inf_{u'\in\dom(u)}(u(u')\Then\val{u'\in v}).
\eeqas
Thus, relation (ii) follows from the definition of $\val{u\subseteq x}$. 
Relation (iii) follows easily from relation (ii).
\end{proof}

\begin{thm}[$\De_0$-Restriction Principle]
\label{th:V-restriction}
For any $\De_{0}$-formula ${\vp} (x_{1},{\ldots}, x_{n})$ 
in $\LL_{\in}$ and $u_{1},{\ldots}, u_{n}\in\VQ$, if 
$p\in S(u_1,\ldots,u_n)^{!}$, then 
$\val{\vp(u_1,\ldots,u_n)}\And p=
\val{\vp(u_1|_p,\ldots,u_n|_p)}\And p$.
\end{thm}
\begin{proof}
We prove the assertion by induction on 
the complexity of  ${\vp} (x_{1},{\ldots},x_{n})$.
From Proposition \ref{th:restriction-atom}, the assertion
holds for atomic formulas.
Then the verification of every induction step follows 
from the fact that (i) the relation
$a^{\perp}\And p=(a\And p)^{\perp}\And p$ holds
for  all $a,b\in\{p\}^{!}$,
(ii) 
the relation $(a\Then b)\And p=[(a\And p)\Then (b\And p)] 
\And p$ holds for  all $a,b\in\{p\}^{!}$
from property (I2) of the generalized implication $\Then$,
(iii) 
the function $a\mapsto a\And p$
of all $a\in\{p\}^{!}$ preserves the supremum and the infimum
as shown in Proposition \ref{th:logic}, and that
(iv) 
the generalized implication satisfies 
relation (ii) of Theorem \ref{th:implication}.
\end{proof}

Let $\cU\subseteq\VQ$.  The {\em commutator of $\cU$}, denoted by
$\com(\cU)$, is defined by 
$$
\com(\cU)=\com (S(\cU)).
$$
For any $u_1,\ldots,u_n\in\VQ$, we write
$\com(u_1,\ldots,u_n)=\com(\{u_1,\ldots,u_n\})$ and 
 $\com(\vec{u})=\com(u_1,\ldots,u_n)$ if $\vec{u}=(u_1,\ldots,u_n)$.

Now, we can prove the following.

\begin{thm}[Quantum Transfer Principle]\label{th:QTP}
For any $\De_{0}$-formula ${\vp} (x_{1},{\ldots}, x_{n})$ 
of $\LL_{\in}$ and $u_{1},{\ldots}, u_{n}\in\VQ$, if 
${\vp} (x_{1},{\ldots}, x_{n})$ is provable in ZFC, then
we have
$$ 
\val{\vp({u}_{1},\ldots,{u}_{n})}\ge\com(u_{1},\ldots,u_{n}).
$$
\end{thm}
\begin{proof}
Let $p=\com(u_1,\ldots,u_n)$.
Then  $a\And p\commutes b\And p$
for any $a,b\in S(u_1,\ldots,u_n)$, and hence 
there is a Boolean sublogic $\cB$ such that 
$S(u_1,\ldots,u_n)\And p\subseteq \cB$.
From Proposition \ref{th:L-restriction},
we have $S(u_1|_p,\ldots,u_n|_p)\subseteq \cB$.
From Proposition \ref{th:sublogic}, we have
$u_1|_p,\ldots,u_n|_p\in \VB$.
By the Boolean Transfer Principle (Theorem \ref{th:BTP}),
we have
$\val{\vp(u_1|_p,\ldots,u_n|_p)}{}_{\cB}=1$. By the
$\De_0$-absoluteness principle, we have
$\val{\vp(u_1|_p,\ldots,u_n|_p)}=1$.
From Proposition \ref{th:V-restriction}, we have
$\val{\vp(u_1,\ldots,u_n)}\And p
=\val{\vp(u_1|_p,\ldots,u_n|_p)}\And p
=p$, and the assertion follows.
\end{proof}

From the Boolean Transfer Principle  (Theorem \ref{th:BTP}) 
if the logic $\cQ$ is a Boolean algebra,
\[
\val{\vp({u}_{1},\ldots,{u}_{n})}=1
\]
holds for any formula ${\vp} (x_{1},{\ldots}, x_{n})$ in $\LL_{\in}$ provable in ZFC.  
We also obtain the converse of the Boolean
Transfer Principle.

\begin{thm}[Converse of the Boolean Transfer Principle]
If the relation
\[
\val{\vp({u}_{1},\ldots,{u}_{n})}=1
\]
holds 
for any formula ${\vp} (x_{1},{\ldots}, x_{n})$ in $\LL_{\in}$ provable in ZFC
and $u_1,\ldots,u_n\in\VQ$ then $\cQ$ is a Boolean algebra.
\end{thm}
\begin{proof}
Let $P,Q\in\cQ$.
Define $\tP,\tQ\in\VQ$ by
$\tP=\bracket{\ck{0},P}$ and $\tQ=\bracket{\ck{0},Q}$,
i.e., 
$\dom(\tP)=\dom(\tQ)=\{\check{0}\}$ and 
$\tP(\check{0})=P$ and $\tQ(\check{0})=Q$.
Then by definition we have
$\val{\check{0}\in\tP}=P$, 
 $\val{\check{0}\not\in\tP}=P^{\perp}$,
$\val{\check{0}\in\tQ}=Q$,
 and $\val{\check{0}\not\in\tQ}=Q^{\perp}$.
Note that the above relations hold independentl of the choice of the generalized implication
$\Then$ in $\cQ$. 
Since the formula
\[
z\in x \IFF [(z\in x \And z\in y)\Or(z\in x \And z\not\in y)]
\] 
is provable in ZFC, where the connective $\IFF$ is defined by 
\[
\vp\IFF\ps := (\vp\And \ps)\Or(\Not\vp\And\Not\ps),
\]
by assumption we have
\[
\val{\ \check{0}\in \tP\IFF [(\check{0}\in \tP 
\And \check{0}\in \tQ)\Or(\check{0}\in \tP 
\And \check{0}\not\in \tQ)]\, }=1.
\]
Thus, we obtain
\[
\left(
\val{\check{0}\in \tP}\IFF (\val{\check{0}\in \tP} 
\And \val{\check{0}\in \tQ})\Or(\val{\check{0}\in \tP} 
\And \val{\check{0}\not\in \tQ})\right)=1,
\]
where  the operation  $\IFF$  on $\cQ$ is defined by 
$X\IFF Y=(X\And Y)\Or(X^{\perp}\And Y^{\perp})$
for all $X,Y\in\cQ$.
Therefore, the relation 
$P=(P\And Q)\Or(P\And Q^{\perp})$ follows, and we conclude $P\commutes Q$.
Since  $P,Q\in\cQ$ were arbitrary, we conclude that $\cQ$ is a Boolean algebra. 
\end{proof}

In our definition of  $(\cQ,\Then)$-valued interpretation, 
we assumed that $\Then$ was one of the generalized implications.
Now we extend the definition to arbitrary binary operations $\Then$ on $\cQ$.
Then Theorem \ref{th:QTP} shows that if $\Then$ is a generalized
implication then the Quantum Transfer Principle holds for the 
$(\cQ,\Then)$-valued interpretation.
We conclude this paper by asking which binary polynomials $\Then$ on $\cQ$
allow the Quantum Transfer Principle for the $(\cQ,\Then)$-valued interpretation:
the six polynomially definable generalized implications do so, and no others.

\begin{thm}
Let $(\cQ,\Then)$ be a logic with
an arbitrary binary operation $\Then$ on $\cQ$. 
Suppose that the $(\cQ,\Then)$-interpretation of 
$\VQ$ satisfies the Quantum Transfer Principle, i.e., 
$$ 
\val{\vp({u}_{1},\ldots,{u}_{n})}\ge\com(u_{1},\ldots,u_{n})
$$
holds in the $(\cQ,\Then)$-interpretation 
for any $\De_{0}$-formula ${\vp} (x_{1},{\ldots}, x_{n})$ 
of $\LL_{\in}$ provable in ZFC and $u_{1},{\ldots}, u_{n}\in\VQ$.
Then the operation $\Then$ satisfies (LB).
In particular, the polynomially definable binary operations $\Then$ 
with which the Quantum Transfer Principle holds
for the $(\cQ,\Then)$ interpretation
are exactly the six operations
$\Then_j$ for $j=0,\ldots,5$.
\end{thm}

\begin{proof}
Suppose that $P,Q\in\cQ$ and $P\commutes Q$ or equivalently $\com(P,Q)=1$.
Let $\tP=\bracket{\ck{0},P}$ and $\tQ=\bracket{\ck{0},Q}$. 
Let $\vp(x_1,x_2,x_3)$ be the $\De_0$-formula in $\LL_{\in}$ such that
$$
\vp(x_1,x_2,x_3):=(x_1\in x_2\Then x_1\in x_3)
\IFF \left(\Not(x_1\in x_2)\Or(x_1\in x_3)\right),
$$
where the connective $\IFF$ is defined by 
$X\IFF Y:=(X\And Y)\Or(X^{\perp}\And Y^{\perp})$.
Then  $\vp(x_1,x_2,x_3)$ is a tautology in classical logic 
and a theorem of ZFC set theory.
We have $\com(\ck{0},\tP,\tQ)=\com(P,Q)=1$.  
By the Quantum Transfer Principle, we have 
$\val{\vp(\ck{0},\tP,\tQ)}\ge\com(\ck{0},\tP,\tQ)=1$.
Thus, we have
\beqas
\val{\ck{0}\in \tP\Then \ck{0}\in\tQ}=\val{\Not(\ck{0}\in\tP)\Or(\ck{0}\in\tQ)}.
\eeqas
Since we have 
$\val{\check{0}\in\tP}=P$, 
 $\val{\Not(\check{0}\in\tP)}=P^{\perp}$,
$\val{\check{0}\in\tQ}=Q$,
 and $\val{\Not(\check{0}\in\tQ)}=Q^{\perp}$,
 we conclude 
 $$
 P\Then Q=P^{\perp}\Or Q.
 $$
 Since $P,Q\in\cQ$ are arbitrary elements with $P\commutes Q$,
the operation $\Then$ satisfies (LB).
Thus, from Theorem \ref{th:QTP} for any binary ortholattice polynomial $P\Then Q$ on $\cQ$,
the $(\cQ,\Then)$-interpretation of $\cQ$ satisfies the Quantum Transfer Principle
if and only if $\Then$ satisfies (LB), and hence the rest of the assertion follows from 
the  characterization of the polynomially definable operations satisfying (LB) due to
\cite{Kot67}.
\end{proof}

\xsection{Concluding remarks: Applications to quantum mechanics.}
\label{se:CR}

In quantum mechanics, every system $\bS$ is described by a Hilbert space $\cH$,
a state of $\bS$ is represented by a vector in $\cH$, 
and an observable of $\bS$ is represented by a self-adjoint operator densely 
defined in $\cH$.
Here, we assume $\dim(\cH)<\infty$ for simplicity; 
see the Appendix for a more general treatment.
For any observable $A$ and any real number $a\in\R$, we introduce an 
{\em observational proposition} $A=a$ meaning that 
``the observable $A$ takes the value $a$''.
Then $A=a$ holds in a state $\ps$ if and only if
$\ps$ is an eigenstate of $A$ belonging to $a$, i.e., $A\ps=a\ps$.
We write $\ps\forces A=a$ if $A\ps=a\ps$ and
define $\valo{A=a}=\cP(\{\ps\in\cH\mid \ps\forces A=a\})$.
Then 
$$
\valo{A=a}=P^{A}(a),
$$
where $P^{A}(a)=\cP(\{\ps\in\cH\mid A\ps=a\ps\})$.
According to the superposition principle, 
we say that $A=a$ holds 
with probability $p$ in the state $\ps\not=0$ if $\ps=\ps'+\ps''$ with
$\ps'\perp\ps''$,  $\ps'\forces A=a$, and $p=\|\ps'\|^2/\|\ps\|^2$,
or equivalently  $p=\|\,\valo{A=a}\ps\,\|^{2}/\|\ps\|^2$.
Thus, $A=a$ does not hold in $\ps$ if and only if
$\ps\perp\ps'$ for any $\ps'\in\cH$ such that $\ps'\forces A=a$.
We introduce negation as $\ps\forces\Not (A=a)$ if and only if 
$\ps\perp\ps'$ for any $\ps'\in\cH$ such that $\ps'\forces A=a$.
We define $\valo{\Not(A=a)}=\cP(\{\ps\in\cH\mid \ps\forces\Not (A=a)\})$.
Then  
$$
\valo{\Not (A=a)}=\valo{A=a}^{\perp}.
$$
For two observables $A$ and $B$, the observable $A$ takes the value
$a\in\R$ and {\em simultaneously} the observable $B$ takes the value $b\in\R$
in a state $\ps\in\cH$ if and only if the state $\ps$ is a common eigenstate 
of $A$ and $B$ belonging to the respective eigenvalues $a$ and $b$,
i.e., $A\ps=a\ps$ and $B\ps=b\ps$.
We introduce conjunction $\And$ 
by $\ps\forces A=a \And B=b$ if and only if $\ps\forces A=a$ and 
$\ps\forces B=b$. We define $\valo{A=a\And B=b}
=\cP(\{\ps\in\cH\mid \ps\forces A=a\And B=b\})$.
Then 
$$\valo{A=a\And B=b}=\valo{A=a}\And\valo{B=b}.$$
In contrast to the interpretation provided by \cite{BvN36},
we do not require that $A$ and $B$ commute to
introduce conjunction.
We introduce the connective $\Or$ by De Morgan's law, so that
$\ps\forces A=a\Or B=b$ if and only if $\ps\forces
\Not[\Not(A=a)\And\Not(B=b)]$.  We define 
$\valo{A=a\Or B=b}=\{\ps\in\cH\mid\ps\forces A=a\Or B=b\}$.
Then  
$$
\valo{A=a\Or B=b}=\valo{A=a}\Or\valo{B=b}.
$$
We call any formula constructed from observational propositions of the
form $A=a$ with connectives $\Not$, $\And$, and $\Or$ as an observational
proposition.  Then we can define $\valo{\vp}$ for all observational 
propositions by the above relations, since for any observational proposition
$\vp$ there exists an observable $E$ such that $\valo{E=1}=\valo{\vp}$.
In fact, if we have determined $\valo{\vp_1}$ and $\valo{\vp_2}$ 
for two observational propositions $\vp_1$ and $\vp_2$, 
there exist two projections $E_1$ and $E_2$ such that 
$\valo{E_1=1}=\valo{\vp_1}$ and $\valo{E_2=1}=\valo{\vp_2}$.
Thus, the relation 
$$
\valo{\vp_1\And\vp_2}=\valo{\vp_1}\And\valo{\vp_2}
$$ 
is obtained by
$$
\valo{\vp_1\And\vp_2}=\valo{E_1=1\And E_2=1}
=\valo{E_1=1}\And\valo{E_2=1}
=\valo{\vp_1}\And\valo{\vp_2}.
$$
Similarly, we obtain the relations
\beqas
\valo{\Not\vp_1}&=&\valo{\vp_1}^{\perp},\\
\valo{\vp_1\Or\vp_2}&=&\valo{\vp_1}\Or\valo{\vp_2}.
\eeqas
We also determine the probability $\Pr\{\vp\|\ps\}$ of any 
observational proposition $\vp$ in a state $\ps$ as
$$
\Pr\{\vp\|\ps\}=\frac{\|\,\valo{\vp}\ps\,\|^2}{\|\,\ps\,\|^2}
$$
from the relations
$$
\Pr\{\vp\|\ps\}=\Pr\{E=1\|\ps\}=\frac{\|\,\valo{E=1}\ps\,\|^2}{\|\,\ps\,\|^2}
=\frac{\|\,\valo{\vp}\ps\,\|^2}{\|\,\ps\,\|^2},
$$
where the projection $E$ is given by $E=\valo{\vp}$ so that 
$\valo{\vp}=\valo{E=1}$ holds.

\cite{Kot67} showed that any polynomially definable binary operation 
on Boolean algebras has six variations on general orthomodular lattices.
For conjunction we have the following six polynomially
definable binary operations $\And_{j}$ for $j=0,\ldots,5$ on a logic $\cQ$
satisfying $P\And_j Q=P\And Q$  if $P\commutes Q$ for all $P,Q\in\cQ$.
\benum
\item[\rm (i)]
$P\And_0 Q=P\And Q.$
\item[\rm (ii)]
$P\And_1 Q=(P\And_0 Q)\Or (P\And \com(P,Q)\p)$.
\item[\rm (iii)]
$P\And_2 Q=(P\And_0 Q)\Or (Q\And \com(P,Q)\p)$.
\item[\rm (iv)]
$P\And_3 Q=(P\And_0 Q)\Or (P\p\And \com(P,Q)\p)$.
\item[\rm (v)]
$P\And_4 Q=(P\And_0 Q)\Or (Q\p\And \com(P,Q)\p)$.
\item[\rm (vi)]
$P\And_5 Q=(P\And_0 Q)\Or \com(P,Q)\p$.
\eenum

Our choice of $\And_0$ for conjunction is derived from the quantum
mechanical interpretation that $\ps\forces A=a \And B=b$ holds if and only 
if the observable $A$ takes the value $a\in\R$ and simultaneously 
the observable $B$ takes the value $b\in\R$ in the state $\ps\in\cH$.

Similarly for disjunction we have the following six polynomially definable 
binary operations $\Or_j$ for $j=0,\ldots,5$ on a logic $\cQ$ satisfying
$P\Or_j Q=P\Or Q$ if $P\commutes Q$ for all $P,Q\in\cQ$.
\benum
\item[\rm (i)]
$P\Or_0 Q=(P\And Q)\Or(P\And Q^{\perp})\Or(P^{\perp}\And Q).$
\item[\rm (ii)]
$P\Or_1 Q=(P\Or_0 Q)\Or (P\And \com(P,Q)\p)$.
\item[\rm (iii)]
$P\Or_2 Q=(P\Or_0 Q)\Or (Q\And \com(P,Q)\p)$.
\item[\rm (iv)]
$P\Or_3 Q=(P\Or_0 Q)\Or (P\p\And \com(P,Q)\p)$.
\item[\rm (v)]
$P\Or_4 Q=(P\Or_0 Q)\Or (Q\p\And \com(P,Q)\p)$.
\item[\rm (vi)]
$P\Or_5 Q=P\Or Q$.
\eenum

Our choice of $\Or_5$ for disjunction is derived from De Morgan's law,
which makes $\cQ$ a lattice with conjunction and disjunction.

As above, we have naturally derived that the logical structure $\cQ$ 
of observational propositions on a quantum system $\bS$ described by
a Hilbert space $\cH$ forms a complete orthocomplemented
modular (if $\dim(\cH)<\infty$) or orthomodular (if $\dim(\cH)=\infty$)
lattice $\cQ=\cQ(\cH)$  with conjunction, 
disjunction, and negation \citep{BvN36,Hus37}.  
However, there still exists arbitrariness of choosing
the operation for implication from the six polynomially
definable binary operations $\Then_{j}$ for $j=0,\ldots,5$ on the logic $\cQ$
satisfying $P\Then_j Q=P^{\perp}\Or Q$  if $P\commutes Q$ for all $P,Q\in\cQ$
(cf. Theorem \ref{th:poly}).

In this paper, we have shown that for any polynomially definable binary
operation $\Then$ on the orthomodular lattice $\cQ$, 
the Quantum Transfer Principle holds for the $(\cQ,\Then)$ interpretation 
of the language $\LL_{\in}$ of set theory if and only if $\Then$ is one
of the six operations $\Then_{j}$ for $j=0,\ldots,5$.
Thus, quantum set theory can be developed under a very flexible formulation
with a strong logical tool for interpreting theorems of ZFC set
theory.

For further selections among the six polynomially definable generalized
implications, recall that
\cite{Har81} proposed the following requirements for the implication connective.
\benum
\item[(E)]    $P\Then Q=1$ if and only if $P\le Q$ for all $P,Q\in\cQ$.

\item[(MP)] $P             \And (P\Then Q) \le Q$ for all $P,Q\in\cQ$.

\item[(MT)] $Q^{\perp}\And (P\Then Q) \le P^{\perp}$ for all $P,Q\in\cQ$.

\item[(NG)] $P\And Q\p\le(P\Then Q)^\perp$ for all $P,Q\in\cQ$.

\eenum
\cite{Har81} showed that requirement (E) is satisfied by $\Then_j$ for $j=0,\ldots,4$
and that all requirements (E), (MP), (MT), and (NG) are satisfied by
$\Then_j$ for $j=0,2,3$, where 
$\Then_0$ is called the minimum implication or the relevance implication
\citep{Geo79},  $\Then_2$ is called the contrapositive Sasaki arrow,
and  $\Then_3$ is called the Sasaki arrow \citep{Sas54}.

In the previous investigations on quantum set theory only the Sasaki arrow
$\Then_3$
has been studied as the implication connective \citep{Ta81,07TPQ,16A2},
in which the Quantum Transfer Principle has been established,
and also the structure of the real numbers in the model $\VQ$ has been
figured out.
\cite{Ta81} has shown that the real numbers in $\VQ$ are in
one-to-one correspondence with the observables (self-adjoint operators) 
in $\cH$.
In our previous study \citep{07TPQ}, 
the Quantum Transfer Principle for the $(\cQ,\Then_3)$ interpretation 
has been established and it has been shown that equality between real 
numbers in $\VQ$ satisfies the equality axioms.
In the recent study \citep{16A2}, the embedding $\vp\mapsto\widetilde{\vp}$
of the observational propositions into the sentences in $\LL_{\in}(\VQ)$ 
are defined with the embedding $A\mapsto\widetilde{A}$ of the set
$\cO(\cH)$ of observables in $\cH$ onto the set $\R^{(\cQ)}$ of
real numbers in $\VQ$ so that the relations
\beqas
\widetilde{A=a}&=&\widetilde{A}=\widetilde{a},\\
\widetilde{\Not\vp}&=&\Not\widetilde{\vp},\\
\widetilde{\vp_1\And\vp_2}&=&\widetilde{\vp}_1\And\widetilde{\vp}_2,\\
\widetilde{\vp_1\Or\vp_2}&=&\widetilde{\vp}_1\Or\widetilde{\vp}_2,\\
\widetilde{\vp_1\Then\vp_2}&=&\widetilde{\vp}_1\Then\widetilde{\vp}_2,\\
\val{\ph}_0&=&\val{\ph}
\eeqas
hold for all $A\in\cO$ and all observational propositions $\vp$,
and the standard interpretation of quantum mechanics has been
extended to introduce new observational propositions $A=B$ by
$$
\val{A=B}_0=\val{\widetilde{A}=\widetilde{B}}
$$ 
for any $A,B\in\cO(\cH)$, while it has been shown that $\ps\forces A=B$
if and only if $A$ and $B$ are perfectly correlated in $\ps$, or equivalently
$A$ and $B$ commute in $\ps$ and they have the joint probability distribution
$\mu^{A,B}_{\ps}$ concentrating on the diagonal, i.e., 
$\mu^{A,B}_{\ps}(a,b)=0$ if $a\not= b$ for all $a,b\in\R$ \citep{05PCN,06QPC}.

The above close connections between quantum mechanics and real number
theory in $\VQ$ have been obtained for the $(\cQ,\Then_3)$-interpretation.
However, it will be an interesting program to extend the relation between
quantum mechanics and quantum set theory to other interpretations
with other generalized implications $\Then$.
In particular, it will be of particular significance to figure out what generalized
implications allow the isomorphism between observables and 
real numbers in $\VQ$ and whether there arise any operational differences
in extending the interpretation of quantum mechanics using the $(\cQ,\Then)$
interpretation of quantum set theory for different generalized 
implications $\Then$.

\section*{Acknowledgments.}
This work was supported by JSPS KAKENHI, No.~26247016.
The author thanks Minsheng Ying for useful comments on an earlier version.

\appendix

\section*{Appendix. Observational propositions for a quantum system described by
a von Neumann algebra.}

In this section, we consider the logical structure of observational 
propositions of a (local) quantum system $\bS$ 
described by a von Neumann algebra $\cM$ on a Hilbert space $\cH$ \citep{Ara00}.
In this formulation, an observable of the system $\bS$ 
is represented by a self-adjoint operator $A$ densely defined in $\cH$
satisfying $E^{A}(a)\in\cM$ for any $a\in\R$, where $E^{A}(a)$
is the resolution of the identity belonging to $A$ \cite[p.~119]{vN55}.
Denote by $\cO(\cM)$ the set of observables of $\bS$. 
For any $A\in\cO(\cM)$ and $a\in\R$, we introduce an 
{\em observational proposition} $A\le a$ meaning that 
``the observable $A$ takes the value $\le a$''.
Any vector $\ps\in\cH$ represents a state of $\bS$.
Then  $A\le a$ holds in $\ps\in\cH$, in symbols $\ps\forces A\le a$,
if and only if  $\ps\in\ran( E^{A}(a))$.
Define $\valo{A\le a}=\cP(\{\ps\in\cH\mid \ps\forces A\le a\})$.
Then 
$$
\valo{A\le a}=E^{A}(a).
$$
According to the superposition principle, 
$A\le a$ holds with probability $p$ 
and does not hold with probability $1-p$ 
in the state $\ps\not=0$ if and only if $\ps=\ps'+\ps''$ with
$\ps'\perp\ps''$,  $\ps'\forces A\le a$, $p=\|\ps'\|^2/\|\ps\|^2$,
and $1-p=\|\ps''\|^2/\|\ps\|^2$.
Thus, $A\le a$ does not hold in $\ps$ with probability 1 if and only if
$\ps\perp\ps'$ for any $\ps'\in\cH$ such that $\ps'\forces A\le a$.
We introduce negation as $\ps\forces\Not (A\le a)$ if and only if
$\ps\perp\ps'$ for any $\ps'\in\cH$ such that $\ps'\forces A\le a$.
We define $\valo{\Not(A\le a)}=\cP(\{\ps\in\cH\mid \ps\forces\Not (A\le a)\})$.
Then  
$$\valo{\Not (A\le a)}=\valo{A\le a}^{\perp}.$$
For two observables $A$ and $B$, the observable $A$ takes the value
$\le a\in\R$ and {\em simultaneously} the observable $B$ takes the value $\le b\in\R$
in a state $\ps\in\cH$ if and only if the state $\ps$ is a common eigenstate 
of $E^A(a)$ and $E^B(b)$ with eigenvalue 1, or equivalently 
$\ps\forces A\le a$ and $\ps\forces B\le b$.
We introduce conjunction $\And$ 
by $\ps\forces A\le a \And B\le b$ if and only if $\ps\forces A=a$ and 
$\ps\forces B=b$. We define $\valo{A\le a\And B\le b}
=\cP(\{\ps\in\cH\mid \ps\forces A\le a\And B\le b\})$.
Then 
$$\valo{A\le a\And B\le b}=\valo{A\le a}\And\valo{B\le b}.$$
We introduce the connective $\Or$ by De Morgan's law, so that
$\ps\forces A\le a\Or B\le b$ if and only if $\ps\forces
\Not[\Not(A\le a)\And\Not(B\le b)]$.  We define 
$\valo{A\le a\Or B\le b}=\{\ps\in\cH\mid\ps\forces A\le a\Or B\le b\}$.
Then  
$$
\valo{A\le a\Or B\le b}=\valo{A\le a}\Or\valo{B\le b}.
$$
We call any formula constructed from observational propositions of the
form $A\le a$ with connectives $\Not$, $\And$, and $\Or$ an observational
proposition.  Then we can define $\valo{\vp}$ for all observational 
propositions by a method similar to the one given in \ref{se:CR}
to obtain the relations
\beqas
\valo{\Not\vp_1}&=&\valo{\vp_1}^{\perp},\\
\valo{\vp_1\And\vp_2}&=&\valo{\vp_1}\And\valo{\vp_2},\\
\valo{\vp_1\Or\vp_2}&=&\valo{\vp_1}\Or\valo{\vp_2}.
\eeqas
Therefore, the logical structure of observational 
propositions on the system $\bS$
described by a von Neumann algebra $\cM$ 
is also represented by the ortholattice structure of 
the projection lattice $\cP(\cM)$ with the 
interpretations of the logical connectives $\Not$, $\And$, and $\Or$
given above.

\vspace*{10pt}

\noindent
GRADUATE SCHOOL OF INFORMATICS\\
\hspace*{9pt}
NAGOYA UNIVERSITY\\
\hspace*{18pt}
CHIKUSA-KU, NAGOYA,  464-8601, JAPAN\\
{\it E-mail}: ozawa@is.nagoya-u.ac.jp

\clearpage

\end{document}